\documentclass[apj]{emulateapj}
\usepackage{graphicx}
\usepackage{apjfonts}

\newcommand{\cha}{{\it Chandra}}

\shorttitle{Chandra Observations of NGC 4636}
\shortauthors{Posson-Brown et al.}

\submitted{Accepted for publication in ApJ -- January 2009}

\begin{document}

\title{Chandra Observations of the X-ray Point Source Population in
NGC 4636}

\author{Jennifer Posson-Brown\altaffilmark{1}, Somak
  Raychaudhury\altaffilmark{2,1}, William Forman\altaffilmark{1},
  R. Hank Donnelly\altaffilmark{1,3}, Christine
  Jones\altaffilmark{1}}

\altaffiltext{1}{Smithsonian Astrophysical Observatory,
  Harvard-Smithsonian Center for Astrophysics, 60 Garden Street,
  Cambridge, MA 02138}

\altaffiltext{2}{School of Physics and Astronomy, University of
  Birmingham, Edgbaston, Birmingham B15 2TT, United Kingdom}

\altaffiltext{3}{Center for Naval Analysis, 4825 Mark Center Drive,
Alexandria, VA 22311}

\email{jpossonbrown@cfa.harvard.edu} \email{somak@star.sr.bham.ac.uk}
\email{wforman@cfa.harvard.edu} \email{cjones@cfa.harvard.edu}

\begin{abstract}
  We present the X-ray point source population in the nearby Virgo
  elliptical galaxy NGC 4636 from three \cha\ X-ray observations.
  These observations, totaling $\sim$193 ks after time filtering, were
  taken with the Advanced CCD Imaging Camera (ACIS) over a three year
  period. Using a wavelet decomposition detection algorithm, we detect
  318 individual point sources.  For our analysis, we use a subset of
  the 277 detections with $\geq$ net 10 cts (a limiting luminosity of
  approximately $1.2\times10^{37}$ erg~s$^{-1}$ in the 0.5 -- 2 keV
  band, outside the central $1.5^{\prime}$ bright galaxy core).  We
  present a radial distribution of the point sources.  Between 1.5$^\prime$ and 6$^\prime$ from the center, 25\% of our
  sources are likely to be background sources (AGN) and 75\% are LMXBs
  within the galaxy, while at radial distances greater than
  6$^\prime$, background sources (AGN) will dominate the point sources.  We explore short and long-term variability (over time
  scales of 1~day to 3~years) for X-ray point sources in this
  elliptical galaxy.  Fifty-four sources (24\%) in the common ACIS
  fields of view show significant variability between observations.
  Of these, 37 are detected with at least 10 net counts in only one
  observation and thus may be ``transient''.  In addition, $\sim$10\%
  of sources in each observation show significant short term
  variability; we present an example light curve for a variable bright
  source.  The cumulative luminosity function for the point
  sources in NGC 4636 can be represented as a power-law of slope
  $\alpha = 1.14 \pm 0.03$.  We do not detect, but estimate an upper
  limit of $\sim 4.5\times10^{37}$ erg s$^{-1}$ to the
  current X-ray luminosity of the historical supernova SN1939A.  We
  find 77 matches between X-ray point sources and globular cluster
  (GC) candidates found in deep optical images of NGC~4636. In the
  annulus from 1.5$^\prime$ to 6$^\prime$ of the galaxy center, 48 of
  the 129 X-ray point sources (37\%) with $\geq$ 10 net counts are
  matched with GC candidates. Since we
  expect 25\% of these sources to be background AGN, the percentage
  matched with GCs could be as high as 50\%.  Of these matched
  sources, we find that $\sim 70\%$ are associated with the redder GC
  candidates, those that are thought to have near-solar metal
  abundance. The fraction of GC candidates with an X-ray point source
  match decreases with decreasing GC luminosity.  We do not find a
  correlation between the X-ray luminosities of the matched point
  sources and the luminosity or color of the host GC candidates. The
  luminosity functions of the X-ray point sources matched with GCs and those
  that are unmatched have similar slopes over $1.8\times10^{37}$ erg s$^{-1}
\leq L_{x} \leq 1\times10^{38}$ erg s$^{-1}$.  Finally,
  we present a color-color diagram based on ratios of X-ray flux
  rather than source counts, which yields a much tighter source distribution,
  and shows a large population of sources which are likely LMXBs and a
  small population of black hole candidates.

\end{abstract}


\section{Introduction}
\label{s:intro}

The sub-arcsecond resolution of the \cha\ Observatory has revealed the
nature of X-ray emitting point sources in nearby galaxies. It is clear
that almost all the very luminous ($L_{X} > 10^{36}$ erg s$^{-1}$)
point sources in galaxies belong to two distinct populations of
compact binaries, their evolutionary timescales depending upon that of
their donor stars: the low-mass X-ray binaries (LMXB), which are
long-lived and evolve on timescales of $10^9$ -- $10^{10}$~yr, and
high-mass X-ray binaries (HMXB), which evolve on timescales of $10^6$
-- $10^{7}$~yr. The latter population is thus an indicator of recent
star formation and is not expected to be found in early-type galaxies,
unless a recent merger has occurred. The LMXB population, on the other
hand,  has lifetimes comparable with that of the host galaxy, and
their number and combined luminosity is found to correlate well with
the stellar mass of galaxies (Gilfanov 2004).

In this paper, we present a \cha\ view of the X-ray point source
population of NGC~4636, a bright E/S0 galaxy on the southern periphery
of the Virgo cluster. It has a radial velocity similar to that of
Virgo, but is 10.8$^\circ$ from the center of the cluster, which
corresponds to 2.8~Mpc at a distance of 15~Mpc (Tonry et al. 2001).
Furthermore, the galaxy lies at the center of a poor group (Osmond \&
Ponman 2004; Miles et al. 2004, 2005; Baldi et al. 2008), possibly falling into the
cluster.  Its unusual properties have attracted detailed
multiwavelength research for several decades. It has been suggested
that the galaxy has an unusually large dark halo (Loewenstein \&
Mushotzky 2003, Schuberth et al. 2006, Chakrabarty \& Raychaudhury
2008).  NGC~4636 was one of the first early-type galaxies in which
neutral hydrogen was detected (Knapp et al. 1978), and further radio
observations (Birkinshaw \& Davies 1985) revealed a weak central radio
source and small-scale jets.  NGC~4636 has a large population of
globular clusters similar to ellipticals of comparable luminosity
(Dirsch et al. 2005).  Its far-infrared luminosity greatly exceeds
that expected from its stellar content (Temi et al. 2003), and the
luminosity in the vicinity of 100$\mu$ is consistent with dust
emission from a recently accreted disk galaxy. Its flattened (E4)
morphology at the outer faint isophotes (Sandage 1961) indicates the
presence of large-scale angular momentum, often associated with recent
mergers.

NGC~4636 is one the brightest nearby early-type galaxies in X-rays,
and so it has been well-studied with generations of X-ray
observatories.  It was first detected as an extended X-ray source by
\textit{Einstein} (Forman et al. 1985).  ROSAT and ASCA observations
found abundance and temperature gradients in its extended X-ray halo
(Trinchieri et al. 1994; Matsushita et al. 1997, Buote 2000), while
\cha\ and XMM-Newton observations (Jones et al. 2002; O'Sullivan et
al. 2005) show symmetric arm features and cavities in the extended
X-ray halo, interpreted to be evidence of past AGN activity.

The resolution of the \cha\ observatory has enabled the detailed study
of the environment of detected point sources (e.g. Fabbiano 2006 and
references therein). Many of these are associated with globular
clusters.  Since it has been suggested that LMXBs may be primarily
formed in the cores of globular clusters (White et al. 2002), the
properties of point sources, along with those of their host GCs, may
yield important clues to the role played by GCs in the formation of
LMXBs (e.g. Maccarone et al. 2004. Kundu et al. 2007). On the other
hand, there is some evidence that a significant fraction of LMXBs
may be formed in the field, and are thus not expected to be
associated with GCs (Juett 2005, Irwin 2005). The collective study
of X-ray detected point sources and globular cluster candidates
found in optical studies can thus yield important information 
concerning the origin of X-ray emitting binaries in galaxies.

We introduce the observations and discuss data reduction and point
source detection in \S\ref{s:obs}.  In \S\ref{s:lum}, we examine the
distribution of sources in NGC 4636.  
In \S\ref{s:color}, we use soft, medium, and hard
band source fluxes to make an X-ray color-color diagram which is
independent of any instrumental or detector effects.  In
\S\ref{s:var}, we examine the X-ray point source population to look
for variable sources on both long and short timescales.  
In \S\ref{s:lumfun}, we present the luminosity function of
X-ray point sources, and in 
\S\ref{s:opid}, we optically identify sources associated with globular
clusters and examine their properties.  In
\S\ref{s:individ}, we examine a few interesting individual sources.
Finally, our analysis and results are summarized in \S\ref{s:summary}.

\begin{deluxetable}{lllc}
\tablecolumns{4}
\tabletypesize{\small}
\tablecaption{Summary of \cha/ACIS Observations of NGC 4636 \label{t:obsinfo}} 
\tablewidth{0pt}
\tablehead{
\colhead{Date} & \colhead{Detector} & \colhead{Sequence \#} &
\colhead{Exposure Time (s)} 
}
\startdata
2000 Jan 26 & ACIS-S & 600083 & 44450 \\ 
2003 Feb 14 & ACIS-I & 600300 & 74709 \\ 
2003 Feb 15 & ACIS-I & 600331 & 74190 \\ 
\enddata
\end{deluxetable}

\section{Observations, Data Processing, and Point Source Detection}
\label{s:obs}

\begin{figure}[htb!]\center
{\includegraphics[width=3in]{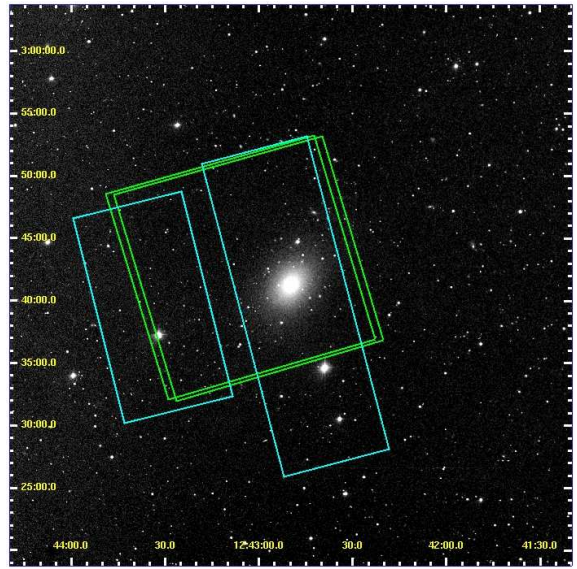}}
\caption{A DSS optical image of NCG 4636 with the FOV of our \cha\
  observations overlaid.  The light blue outline is the January 2000
  ACIS-S observation, and the green outlines are the February 2003
  ACIS-I observations.
\label{f:layout}}
\end{figure}

The observations presented in this paper were made with the ACIS-I and
ACIS-S detectors on \cha\ on three occasions spanning three years.  The
observations are summarized in Table \ref{t:obsinfo}.\footnote{We also
  analyzed a short (5 ks) ACIS-I observation (sequence \#600084) taken
  in December 1999, but it yielded no sources
that were not detected in the other observations, and the source
counts in the few detections present were too low
to allow meaningful analysis.  Thus, we do not use this observation in
the work discussed in the remainder of the paper.}  The FOV of each
observation is overlaid on an optical DSS image in Figure
\ref{f:layout}.

We obtained level 1 event lists for the observations from the \cha\
data archive.  We filtered these event lists to include only the standard good
grade set (ASCA grades 0,2,3,4 and 6) and to exclude bad
pixels.  The event lists were also filtered on the nominal good time
intervals for each observation.  Using the CIAO Contributed S-Lang
Script \textit{lc\_clean.sl}, we did additional time filtering to remove
times affected by background flaring.  We set the script so as to
first calculate an initial mean (3 sigma-clipped), then consider times
with a count rate greater than 1.2 times this mean to be flaring.  The
remaining total good exposure times are listed in Table \ref{t:obsinfo}.

We created a 0.5 -- 2 keV band image for each observation and used the
wavelet decomposition algorithm \textit{wvdecomp} (Vikhlinin et
al. 1995) on the images to detect sources above a threshold of
4.5$\sigma$.  With this detection threshold, and given the image
sizes, we expect approximately 2 false detections per image (Vikhlinin et
al. 1995).  We considered a range of scales from 1 to 6 image pixels
(equivalent to roughly 1 to 6 arcseconds, since 1 ACIS pixel $\approx$
0.5$^{\prime\prime}$ and we binned by a factor of 2).  In addition, we co-added
the two ACIS-I observations done sequentially in February 2003, and
ran the source detection algorithm on this image.  After detection,
source locations were refined using a centroiding algorithm, and the 90\%
encircled energy radius, based on the preflight calibration model of
the \cha\ PSF, was determined for each source location.  We manually
examined the detected sources, looking at them in the ACIS images and
in the individual scale wavelet decomposition images, and rejected
some detections along the galaxy arms which appeared to be knots of
hot gas rather than stellar point sources.  We also rejected spurious
detections along chip gaps.

For each observation, we constructed a 0.5 -- 2 keV band background
image by summing the large-scale (5 and 6) wavelet decompositions with
the residual image (i.e. the image minus the total output of all
scales used).  For each source, we used a background region centered
on the source coordinates with a radius of 1.5 times the 90\% encircled
count fraction radius for the source.  We also made an exposure map for each observation.  For
each source detected in each observation, we tabulated raw and net
(i.e. background-subtracted) counts, and raw and net
exposure-corrected counts.  We compared source positions between the
observations, and matched sources whose coordinates agree within a
dynamic matching radius dependent on the PSFs at the locations of the
sources.  This dynamic matching radius was set to be equal to the
average of the 90\% encircled energy radii of the two sources, or, if this
quantity was smaller than 1.5 arcseconds, the matching radius was set
to 1.5 arcseconds.

As a final step in the initial analysis, we calculated a
counts-to-flux conversion factor for each source by computing response
matrices at each source location on the appropriate ACIS chip,
assuming a power-law model with $\Gamma = 1.5$ (Irwin et al. 2003) and n$_{H} =
1.81\times10^{20}$ cm$^{-2}$ (Dickey \& Lockman 1999).  After comparing fluxes computed with
conversion factors based on different values of $\Gamma$ and n$_{H}$,
we conclude that the choice of parameter values, within a range of
$\Gamma = 0.5 - 2.5$ and n$_{H} = 9.0\times10^{19} - 3.6\times10^{20}$
cm$^{-2}$, changes the resulting fluxes by less than 25\%.  To
convert source fluxes to luminosities, we assume a distance of $d =
15$ Mpc (Tonry et al. 2001).  To facilitate comparison with other
studies, we note that the counts-to-flux conversion factors
for the 0.3-8 keV band are typically 2.2 times greater than those for the 0.5-2
keV band for the ACIS-S, and typically 2.0 times greater for the ACIS-I.

From the 0.5 -- 2 keV background images, we estimate the luminosity a
source would need to be detected at a given location.  By
looking at the maximum values of these ``S$_{min}$'' maps, we estimate
that outside of the bright 1.5$^\prime$ central core, our source
detection is complete above $\sim 1.8\times10^{37}$ erg s$^{-1}$.  Of
the 277 sources that we detect with $\geq$ 10 net counts, 244 are
above this luminosity.

A summary of X-ray point sources detected in the NGC 4636 observations
with at least 10 net counts is given in Table \ref{t:srclist}.

\begin{figure}[htb!]\center
{\includegraphics[width=3in]{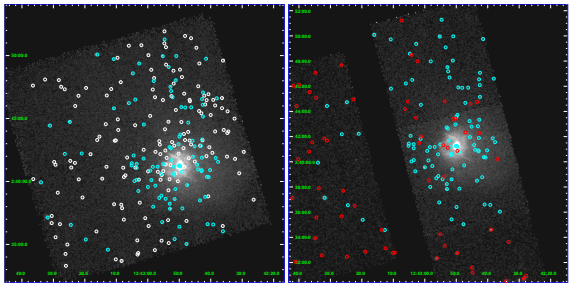}}
\caption{Smoothed, exposure-corrected images in the 0.5 -- 2 keV band.  North is up and East is to the left.  The left image is from
the co-added ACIS-I observations taken in February 2003, totaling
approximately 150 ks.  The right image is from the 45 ks ACIS-S
observation taken in January 2000.  White circles show sources
detected in the ACIS-I image, red circles show sources detected in
the ACIS-S image, and blue circles show sources detected in both
images.  The logarithmic scale (identical in both images) ranges from
0 - 1 $\times$ 10$^{-3}$ cts sec$^{-1}$ pixel$^{-1}$.  
\label{f:img1}}
\end{figure}

\section{Results}
\label{s:results}

\subsection{The radial distribution of the point sources}
\label{s:lum}

\begin{figure}[htb!]\center\rotatebox{90}
{\includegraphics[width=2.5in]{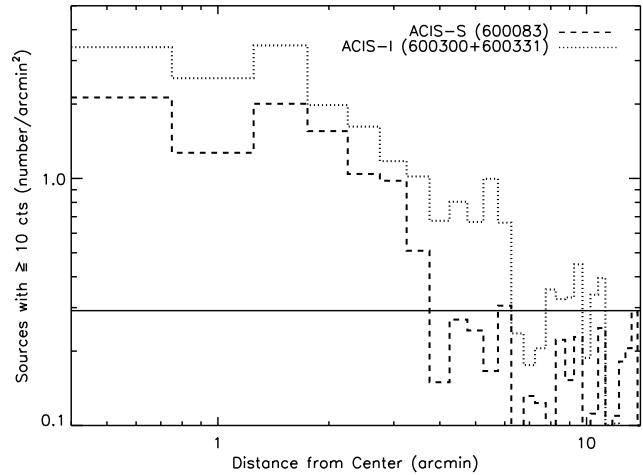}}
\caption{Radial source density profile for sources with $\geq 10$ net
  counts.  The ACIS-S observation (sequence \# 600083) is shown by a
  dashed line and the sequential co-added ACIS-I observations (600300
  and 600331) are shown with a dotted line.   From results of the
  \cha\ Deep Fields (Bauer et al. 2004), we estimate an AGN density of
  $\sim0.3$ arcmin$^{-2}$ at our limiting flux of $\sim4 \times
  10^{-16}$ erg cm$^{-2}$ s$^{-1}$ in the 0.5 -- 2 keV band.  This
  density is shown by the solid line.  Note that our field becomes
  AGN-dominated outside of $\sim6^{\prime}$.
\label{f:surfden}}
\end{figure}

The distribution of the X-ray point sources as a function of
distance from the center of NGC 4636 is shown in
Figure~\ref{f:surfden}.  The radial profiles of sources in the ACIS-S and ACIS-I observations
are shown separately, and are seen to 
have the same shape, with the deeper ACIS-I observations
having a larger number of sources with greater than 10 net counts.

From point source statistics in the \cha\ Deep Fields (Bauer et
al. 2004), we estimate a background AGN density of $\sim 0.3$
arcmin$^{-2}$ at our limiting flux of $\sim 4 \times 10^{-16}$ erg
cm$^{-2}$ s$^{-1}$ in the 0.5 -- 2 keV band (shown as a horizontal
line in Figure~\ref{f:surfden}).  Based on this estimate, our field becomes
AGN-dominated beyond $\sim6^{\prime}$ of the galaxy center.  Between
1.5$^\prime$ and 6$^\prime$, we estimate that 25\% of our sources are
background sources (AGN) and 75\% are LMXBs within the galaxy.

\subsection{Spectral Analysis}
\label{s:color}

\begin{figure}[htb!]\center\rotatebox{90}
{\includegraphics[width=2.5in]{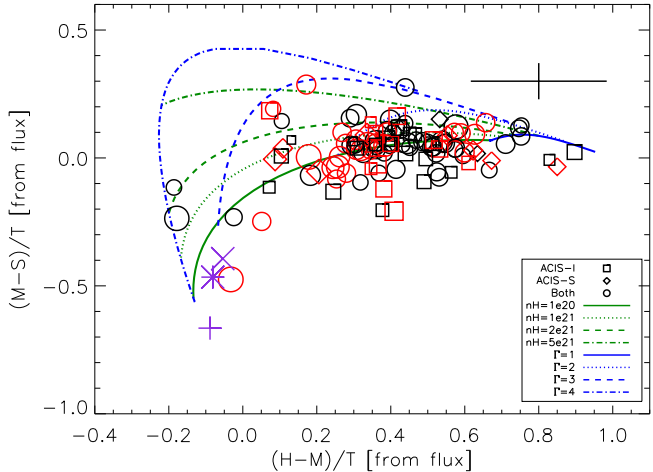}}
\caption{Color-color diagram of all point sources in the NGC 4636
observations having at least 5 net counts in each of the three
bands. We follow the example of Swartz et al. (2004) and use bands S =
0.5 -- 1 keV, M = 1 -- 2 keV, and H = 2 -- 8 keV (T = S+M+H, i.e. 0.5
-- 8 keV).  Note, however, that these color ratios are calculated from
source flux, as opposed to counts.  This allows us to compare
ACIS-I and ACIS-S directly.  The different plotting symbols indicate
whether the source was detected in the ACIS-I observation (open
square), the ACIS-S observation (open diamond) or both observations
(open circle).  The three purple points denote sources detected within
5$^{\prime\prime}$ of the galactic center: the \textit{X} is at the galaxy
nucleus (see \S\ref{s:nucleus}) and the \textit{+} and \textit{*} are two nearby
blackhole candidates (see \S\ref{s:blackhole}).   Plotting symbol size is proportional to source luminosity in the 0.5
-- 2 keV band.  Sources in red lie further than 6$^{\prime}$ from the
center of the galaxy.  The red point near the asterisk is a bursting
X-ray source, discussed in \S\ref{s:burst}.  Blue curves denote colors of absorbed
power-law models of spectral indices $\Gamma$ = 1, 2, 3, and 4 over
the range of absorbing column densities $n_{H} = 10^{20}$ to $10^{24}$
cm$^{-2}$.  Green curves denote constant absorption column densities, as given
in the legend.  A typical error bar is shown in the upper right corner.
\label{f:colcol}}
\end{figure}

Traditionally, X-ray colors for sources are calculated as ratios of counts in
different energy bands (e.g. Swartz et al. 2004; Prestwich et
al. 2003).  However, since we want to directly compare sources
observed by X-ray detectors with different responses (i.e. the ACIS-S
and ACIS-I) and on different parts of the detectors, we convert source
counts to fluxes before computing X-ray colors.  We define 
three bands: soft
(S = 0.5 -- 1 keV), medium (M = 1 -- 2 keV), and hard (H = 2 -- 8
keV), and calculate  two 
colors as C$_{1}$ = (M-S)/(S+M+H) 
and C$_{2}$ = (H-M)/(S+M+H), as defined in Swartz et al. (2004).  The
resulting color-color diagram is shown in Figure~\ref{f:colcol}.  The
plotting symbol size is proportional to the 0.5 -- 2 keV band
luminosity.  We identify two distinct populations: a large cluster of
harder, less luminous sources with power-law indices between 1 and 2,
and a smaller group of softer, more luminous sources ($L_{X} \approx
10^{38}$ erg s$^{-1}$) with steeper power-law indices.  Points in
black lie less than 6$^{\prime}$ from the center of the galaxy and are
most likely members, while points in red lie further than 6$^{\prime}$
from the center.   The source represented by a purple X is
located at the galactic nucleus, and the two sources with steep
spectra represented by the purple asterisk and plus sign lie 
within 5$^{\prime\prime}$
of the nucleus.  These may be signatures of black holes (see
\S\ref{s:nucleus} and \S\ref{s:blackhole}).  The large red circle near
these points is a bursting X-ray source, which we discuss further in
\S\ref{s:burst}.  Due to its distance from the galactic center ($>
11^{\prime}$) it is most likely not a member of NGC 4636.

\begin{figure}[htb!]\center\rotatebox{90}
{\includegraphics[width=2.5in]{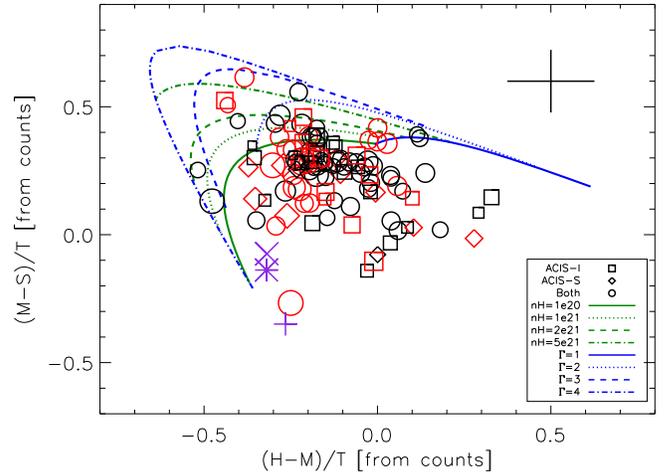}}
\caption{As in Figure \ref{f:colcol}, only here the colors and model curves are
  calculated from counts, whereas in Figure \ref{f:colcol} they are
  calculated from fluxes.  The model curves in this plot are based on the ACIS-I
  response.  Note that here, unlike in Figure
  \ref{f:colcol}, it is not possible to easily group the sources into
  two populations.  
\label{f:colcomp}}
\end{figure}

\begin{figure}[htb!]\center\rotatebox{90}
{\includegraphics[width=2.5in]{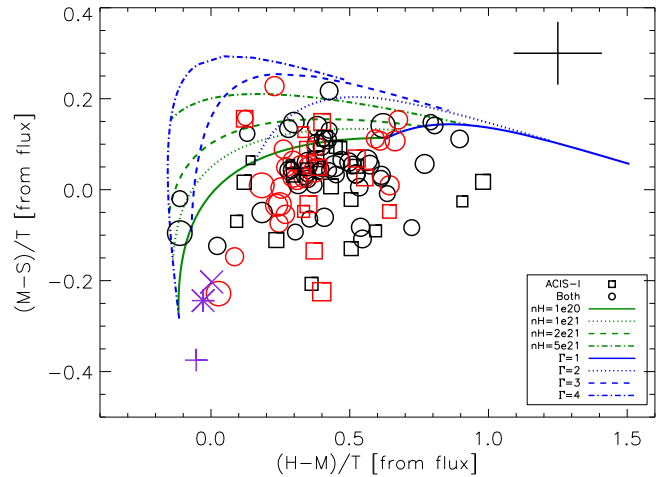}}
\caption{This plot shows the middle step in our iterative flux
  calculation process.  Here, the source and model counts in each band
  have been converted from counts to flux assuming a power-law model with $\Gamma
  = 1.5$ and n$_{H} = 1.81\times10^{20}$ cm$^{-2}$.  This plot shows
  model curves for the ACIS-I, and we have plotted only sources
  detected in the co-added ACIS-I observations.  Based on the location of
  each source in color-color space, we select the nearest model
  curves, and use that  $\Gamma$ and n$_{H}$ to recalculate the
  counts-to-flux conversion factor for that source.  (An identical
  process is followed for sources detected in the ACIS-S observation,
  using model curves for the ACIS-S.)  The results of this iterative
  flux calculation are shown in Figure \ref{f:colcol}.
\label{f:coliter}}
\end{figure}

To make this flux based color-color diagram, we performed an iterative flux
calculation as follows.  Beginning with source counts in
each band for both the ACIS-I and ACIS-S
detectors, we convert to source and model
flux, assuming a power-law model with slope
$\Gamma = 1.5$ and column density n$_{H} =
1.81\times10^{20}$ cm$^{-2}$ and computing response matrices at each
location on the appropriate ACIS chip (as described in \S\ref{s:obs}).
 We convert the model curves from counts to flux assuming the same
 power-law model and using response matrices calculated at the
 aimpoint location of each observation.  Since in earlier works the
 color-color diagram has been based on counts, we show in Figure
 \ref{f:colcomp} the counts-based diagram (with model curves for the
 ACIS-I).  We show in Figure \ref{f:coliter} the color-color diagram from this first flux iteration.
Next, we use model grids and source colors from this first iteration
to determine the values of
$\Gamma$ and n$_{H}$ closest to each source in
color-color space.  We then use these individual  
values of $\Gamma$ and
n$_{H}$ to calculate a more accurate counts-to-flux conversion factor
for the given source (again computing response matrices at each
location on the appropriate ACIS chip).  Note that the model grid shown in Figure \ref{f:coliter} is
a subset of the much finer grid that we use to pick the individual $\Gamma = 1.5$ and
n$_{H}$ for each source.  For example, our grid has $\Gamma$ ranging
from 1 to 4 in steps of 0.1, but for clarity on the plot, we only show
integer values of $\Gamma$.

 After performing this iterative flux
conversion for the sources, we again calculate the colors, now
comparing them to grids 
directly calculated from model flux values.  This is
shown in Figure \ref{f:colcol}.  We note that this figure has much
less scatter than the color-color plot
based on counts for the same set of sources (Figure \ref{f:colcomp}), making it easier to
identify populations and trends.

\begin{figure}[htb!]\center\rotatebox{90}
{\includegraphics[width=2.5in]{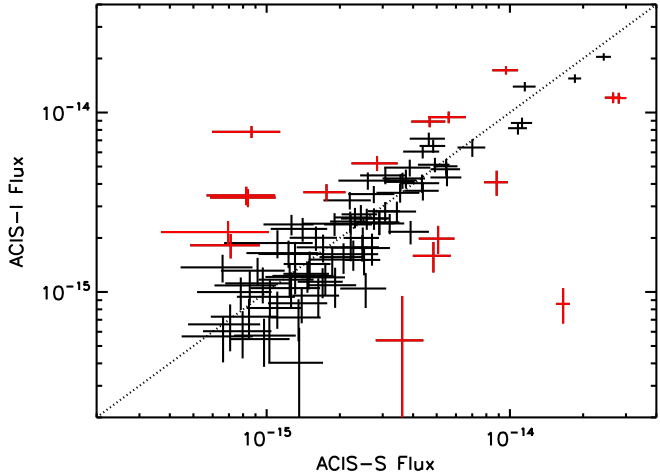}}
\caption{A comparison of source fluxes in the 0.5--2 keV band from the
  45 ks ACIS-S observation and co-added February 2003 ACIS-I
  observations.  Only sources with $\geq$ 10 net counts in both
  observations are plotted here.  Variable sources are shown in red.
\label{f:diagSI}}
\end{figure}

\subsection{Temporal Variability}
\label{s:var}

We present a comparison of source fluxes between the January 2000
ACIS-S observation and co-added February 2003 ACIS-I observations in
Figure \ref{f:diagSI}.  To search for long-term variable sources, we
determine a significance threshold as
\begin{equation}
S = \frac{\left| f_{S} - f_{I}\right|}{\sqrt{\sigma_{I}^{2} +
    \sigma_{S}^{2}}} > 3
\end{equation}
where $f$ is the 0.5 - 2 keV band flux and $\sigma$ is the flux
uncertainty (dominated by the Poisson error on the number of counts --
the error on the counts-to-flux conversion factor is minimal and we do
not include it here).  For the 228 sources in the common field of view
detected with $\geq$ 10 net counts in at least one of the
observations, we find 54 ($\sim$24\%) are long-term variable sources.
These are marked with a ``V'' in the last column of Table
\ref{t:srclist}.  Of these 54 sources, 17 have $\geq$ 10 net counts in
both observations.  These are shown in red in Figure
\ref{f:diagSI}. The remaining 37 variable sources are ``transient'' -- that is, they
have $\geq$ 10 net counts in only one observation, and are not
``reliably'' detected in the other observation.  Of these 37 transient
sources, 31 have $\geq$ 20 net counts in one observation and $<$ 10
net counts in the other observation.

\begin{figure}[htb!]\center\rotatebox{90}
{\includegraphics[width=2.5in]{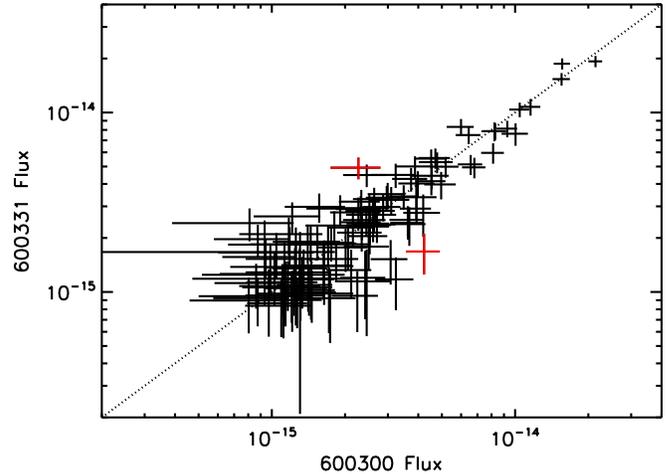}}
\caption{A comparison of source fluxes in the 0.5 -- 2 keV band from
  the 2003 February 14 75 ks ACIS-I observation (sequence number
  600300) and the 2003 February 15 75 ks ACIS-I observation (600331).
  Only sources with $\geq$ 10 net counts in both observations are plotted
  here.  Variable sources are shown in red. 
\label{f:diag03F}}
\end{figure}

To measure variability on shorter timescales, we compare source fluxes
between the two February 2003 ACIS-I observations (see Figure
\ref{f:diag03F}).  These observations were done successively, each
lasting for $\sim$75 ks ($\sim$21 hrs).  Of 188 sources detected
in the common field of view with at least 10 net counts in one
observation or the other, 9 (5\%) vary significantly ($S >$ 3) between
the two observations.  These sources are marked ``MV'' in the last
column of Table \ref{t:srclist}.  Figure \ref{f:diag03F} shows a
subset of these sources: those detected with at least 10 net counts in
both observations.

For yet another look at short term variability, we use the IDL PINTofALE program
\textit{timevarvk} (Kashyap \& Drake 2000), to search for sources
which vary within the course of an observation.  The
\textit{timevarvk} routine uses a one sample
Kolmogorov-Smirnov test to compare the source lightcurve to a flat
model, then calibrates the observed deviation with Monte Carlo
simulations.  Good time intervals are accounted for and gaps are removed. In addition, the deviations between the source
lightcurve and model are averaged over a given number of photons to
suppress random Poisson deviations.  We ran 1000 Monte Carlo
simulations per source, and averaged over 2 photons.  Of 132 sources detected in the ACIS-S
observation with at least 10 net counts, 17 (13\%) are variable at a 3-sigma
level ($p \leq$ 0.0027).  Of 233 sources detected in the summed
ACIS-I observation with at least 10 net counts, 19 (8\%) are variable
at a 3-sigma level.  These sources are marked with ``SV-I'' and/or
``SV-S'' in the last column of Table \ref{t:srclist}.

\begin{figure*}[htb!]\center\rotatebox{90}
{\includegraphics[width=4.5in]{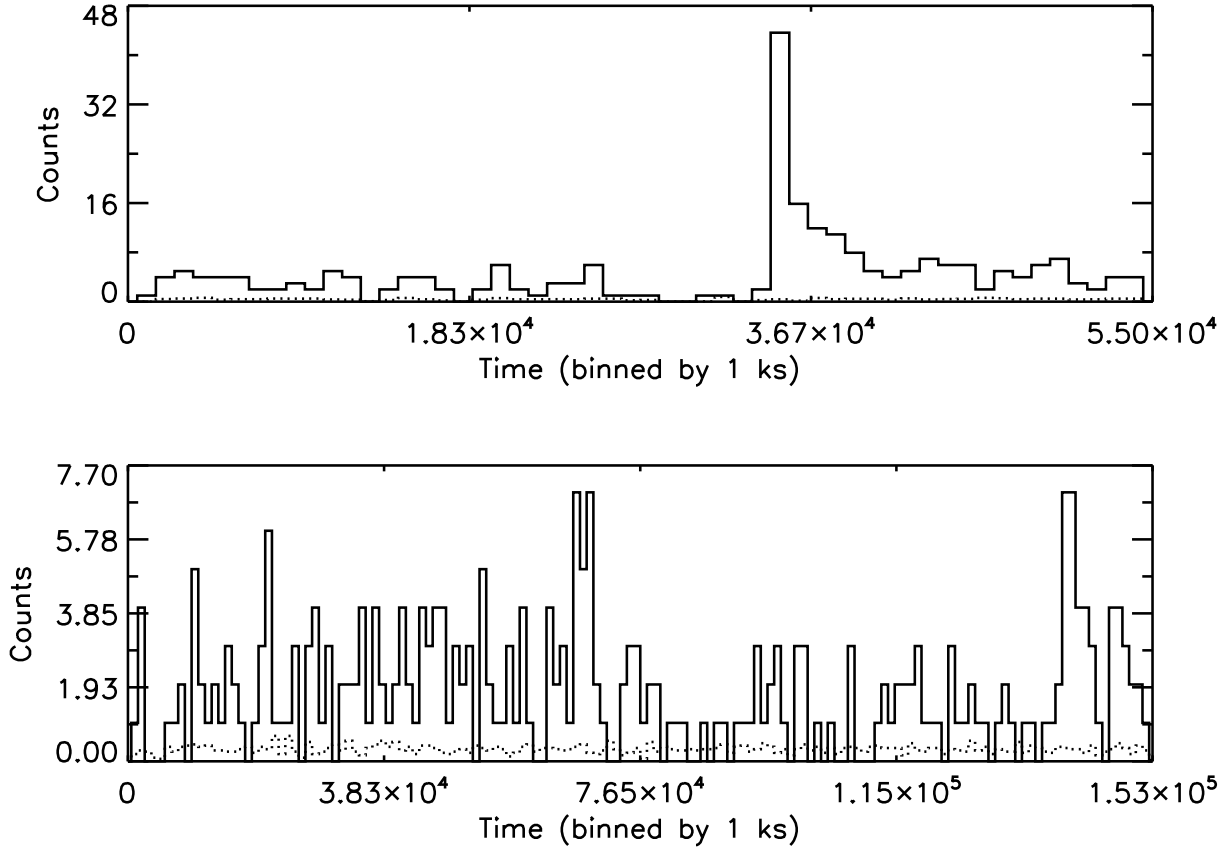}}
\caption{Variable source detected at $\alpha(J2000) =
  190.8825^{\circ}$, $\delta(J2000) = 2.6190^{\circ}$.  This source
  shows both short term (in both observations) and long term (between
  observations) variability.  Counts are in
  the 0.5 -- 2 keV band.  Background counts are plotted with a dotted
  line.  The upper panel is from the ACIS-S observation (sequence \#
  600083) and the lower panel is from the co-added ACIS-I observations
  (sequence \#s 600300 and 600331).  This source is discussed in
  \S\ref{s:burst}.
\label{f:var1}}
\end{figure*}

An example of a variable lightcurve is shown in Figure \ref{f:var1}.
This source shows short term variability during both the ACIS-I and
ACIS-S observations, and also shows variability between the two observations.  Note the flare in the ACIS-S
observation (top panel).  We estimate its duration to be approximately
600 s.  We discuss this bursting X-ray source further in \S\ref{s:burst}.

Comparing our short and long term variability results to X-ray point
source surveys of other galaxies is challenging due to the different
methods and criteria used to determine and quantify variability in each case.  To minimize the effects of
these differences, we attempt to compare  results for a subset of the
brightest sources in each galaxy.  For example, Kraft et al. (2000)
find 35 out of 246 (14\%) of X-ray point sources in Cen
A to be variable with a $\geq$ 3$\sigma$ significance over a 5 month
period separating the two observations.  Computing 
the chi-square statistic for
individual light curves (binsize $\sim$ 3600 s), they find only 2
sources with short term (within an observation - each about 36 ks)
variability.  Of the 17 sources with luminosity greater than $10^{38}$
erg~s$^{-1}$, 8 (47\%) show significant long term variability.  
Jord{\'a}n et al. (2004) search for long term variability among a
subset of sources detected in two M87 observations spanning two
years.  Of the 23 sources with luminosity greater than $10^{38}$
erg~s$^{-1}$, 7 (30\%) vary significantly.  In NGC
4636, we find that of the 52 sources with
luminosity greater than $10^{38}$ erg~s$^{-1}$, 21 (42\%) show some
type of variability: 13 (24\%) long term and 12 (23\%) short term.
(Four of these very luminous sources show both long and short
term variability.)

Loewenstein et al. (2005) find 5 transient sources out of 39 (13\%) and an
additional 6 (15\%) highly variable sources in two observations of NGC
1399 spanning three years.  Three of these 6 sources have luminosities greater
than $10^{39}$ erg~s$^{-1}$, although since luminosities are not
given for the whole source list, we have no basis for comparison
with our results.

Of 126 sources with luminosity greater than 1.4 $\times$
$10^{37}$ erg~s$^{-1}$ in NGC 4697, Sivakoff et al. (2008) find that 5 sources
(4\%) display short term variabilty and 16\% display long term
variability.  Restricting our NGC 4636 population to the 157 sources
lying less than 6$^{\prime}$ from the galactic center with luminosity greater than 1.4 $\times$
$10^{37}$ erg~s$^{-1}$, we find 16 (10\%) that display short term
variability and 29 (18\%) that show long term variability.

\begin{figure*}[htb!]\center\rotatebox{270}
{\includegraphics[width=4in]{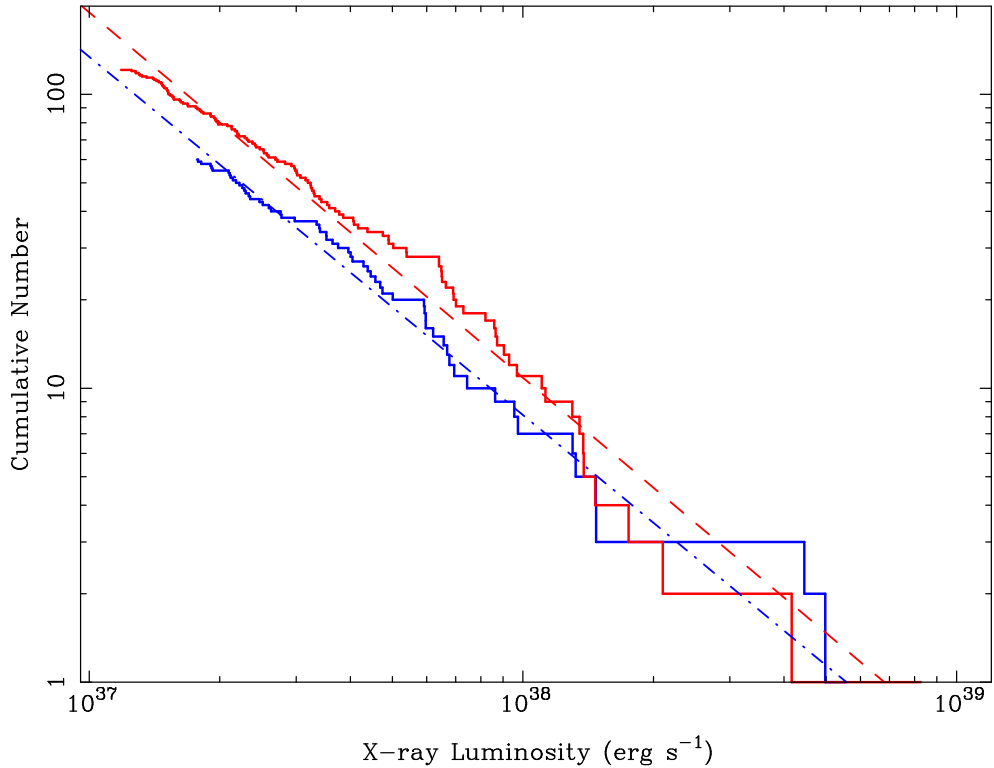}}
\caption{Luminosity function of X-ray point sources in NGC 4636
  detected in the 0.5 -- 2 keV band.  
Only sources with $\geq 10$ net counts 
and radial distance $1.5^{\prime} < d < 6^{\prime}$ from the center
  of the galaxy are included. 
 The ACIS-I data (above, red histogram) are best fit 
(see \S~\ref{s:lumfun}) by a
  power-law with $\alpha$ = 1.14 $\pm$ 0.03 
(dashed line) and the ACIS-S data (below, blue histogram) are
  best fit by a power-law with 
$\alpha$ = 1.19 $\pm$ 0.03 (dot-dashed line). There are no ULXs
in these observations.
  \label{f:NSlum}}
\end{figure*}

\subsection{Luminosity function of point sources}
\label{s:lumfun}

Figure~\ref{f:NSlum} shows the luminosity function in the 0.5 -- 2 keV
band for X-ray point sources 
lying between 1.5$^{\prime}$
and 6$^{\prime}$ from the center of  NGC~4636.  We plot here only the
sources that have 10 net counts, after subtraction of the local
background, in each observation. 
Our requirement 
of having at least 10 net counts per source, leads to a
different minimum luminosity for the two sets of observations
($1.2\times 10^{37}$ erg~s$^{-1}$ and $1.8\times 10^{37}$ erg~s$^{-1}$
for ACIS-I and ACIS-S respectively).
At the bright end, consistent
with Irwin, Bregman \& Athey (2004) and Raychaudhury et al. (2008), we
do not find any ULXs ($L_{X} >$ 1$\times$10$^{39}$ erg~s$^{-1}$).

We model the luminosity function of these sources as a single
power-law, since we see no evidence for a break in the luminosity
function.  Jord\'{a}n et al. (2004) demonstrate that there
is no compelling evidence for a break in the luminosity functions of
the Virgo cluster galaxies M87, M49, and NGC 4697.  We adopt
a robust method of measuring the slope of the luminosity function,
which is usually expressed in the cumulative form
\begin{equation}
\log N(>S) = -\alpha \log S + \kappa.
\label{eq:cumlogn}
\end{equation}
Each of our $n$ point sources has a measured luminosity $F_i$,
given the adopted distance to the galaxy, and an estimated error
$\sigma_i$.  The probability of a point source to have 
a luminosity $S$ is
\begin{equation}
P(S) \,dS = A S^{-\beta}\, dS.
\label{eq:difflogn}
\end{equation}
On comparison with (\ref{eq:cumlogn}), $\beta=1+\alpha$. 
We maximize the log likelihood function
\begin{equation}
\mathcal{L}= \sum^n_{i=1} \ln P(F_i,\sigma_i),
\label{eq:loglike}
\end{equation}
where the distribution
of our 
measured values of flux and error $(F_i, \sigma_i)$
is given by
\begin{equation}
P(F_i,\sigma_i) = \frac{\int_0^\infty P(F_i,\sigma_i | S)\> P(S) 
        \,dS}{\int_{F_{\rm min}}^\infty 
         \int_0^\infty P(F_i,\sigma_i | S)\> P(S) \,dS\,dF}.
\label{eq:probf}
\end{equation}
A more detailed account of this method, and its variants, can
be found in Temple et al. (2005).

Assuming the errors in measuring flux and luminosity are distributed
as a Gaussian, and that the minimum measured flux $F_{\rm min}$ is the
flux of the faintest source (as quoted above) in each observation, we
numerically find the value of $\beta$ (thus $\alpha$) for which
$\mathcal{L}$ in (\ref{eq:loglike}) is maximum, individually for both
observations.

The ACIS-I data (red histogram in Figure~\ref{f:NSlum}) gives the slope
of the cumulative luminosity function $\alpha = 1.14 \pm 0.03$, and
the ACIS-S data (blue histogram) yields $\alpha = 1.19 \pm 0.03$,
which are consistent with each other within errors.  From
samples of 4 and 14 early-type galaxies Gilfanov (2004)
and Kim \& Fabbiano (2004), respectively, found point source
luminosity function slopes in the range $\alpha=$0.6--1.2; our values
for NGC~4636 are at the higher end of this range. 

\begin{figure}[htb!]\center\rotatebox{270}
{\includegraphics[width=2.7in]{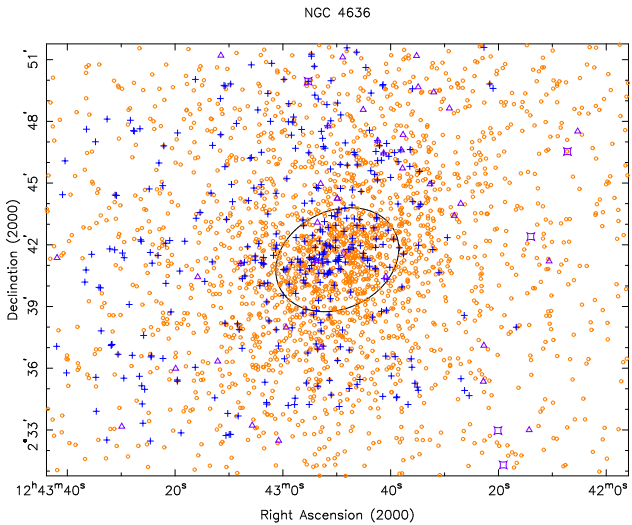}}
\caption{The X-ray point sources detected in NGC~4636, shown by '+'s,
plotted along with all globular cluster candidates (discrete optical
sources from Dirsch et al. (2005), brighter than $R<23.5$ with C-R
color between 0 and 2.5), plotted as
open circles. The optical extent of the galaxy is shown as the
$D_{25}$ ellipse (major and minor axis of 6.0$^{\prime}$ and
4.7$^{\prime}$ respectively). Other known optical sources (from NED)
in the field are also plotted: quasars as squares, and background
galaxies and clusters as triangles. The supernova remnant SN1939A is
plotted as a diamond within the optical extent of the galaxy.
\label{f:skymatch}}
\end{figure}

\subsection{Correlation with the globular cluster population}
\label{s:opid}

It is well-known that a significant fraction of LMXBs in the Milky Way
and elsewhere is associated with globular clusters.  In the Milky Way,
$\sim$10\% of all bright ($>10^{36}$ erg~s$^{-1}$) LMXBs are found in
globular clusters (GCs), even though GCs account for $<10^{-3}$ of the
stellar mass of the galaxy (e.g. Katz 1975, Clark 1975, Grindlay
1993). This has led to the suggestion that LMXBs are very close binary
systems formed as a result of dissipative two-body or three-body
encounters. This could be in the form of encounters between neutron
stars and ordinary stars, which are more likely in the dense cluster
cores (Fabian et al. 1975; Hut et al. 1992), or the exchange of a
companion, where a compact object replaces a member of a binary in a
three-body interaction (Clark 1975, Hills 1976).  Indeed, it has been
suggested that LMXBs are formed primarily in the cores of
globular clusters, and some of the resulting binaries are later
ejected from their host clusters (White et al. 2002). If so, one
expects the LMXB population in a galaxy
to be a good tracer of globular clusters. Alternatively,
a significant fraction of the LMXBs could 
have formed in the field, possibly
as part of the last major star formation episode of the galaxy
(Irwin 2005, Juett 2005), in which case
the populations of GCs and LMXBs need
not be strongly correlated.

In early-type galaxies, \cha\ observations show that there is an
association of LMXBs with globular clusters.  The fraction of LMXBs
identified with known GCs varies from at least 20\% in the Virgo
elliptical NGC~4697
(Sarazin et al. 2001) to up to 70\% in NGC~1399, the central galaxy of
the Fornax cluster (Angelini et al. 2001).  Furthermore, it is
observed that the LMXBs in early-type galaxies are $>$3 times more
likely to be in the redder globular clusters, for galaxies which
exhibit bimodality in color of the GCs, e.g. M87, Jord{\'a}n et
al. 2004; NGC~4472, Kundu et al. 2002; more recently in ten others;
Kundu et al. 2007, Sivakoff et al. 2007; Cen~A (NGC~5128), Minniti et
al. 2004, Jord{\'a}n et al. 2007, Woodley et al. 2008; NGC 3379,
Brassington et al. 2008.  Since the redder GCs are relatively more
metal-rich (e.g. Kundu \& Zepf 2007), this can be a diagnostic of the
characteristics of the compact object's companion star in the LMXB,
and its history of formation (Maccarone et al. 2004; Ivanova
2006). However, this could also result from systematic variation in
the initial mass function of the GCs (Grindlay 1987), a possibility
that has not been observationally explored, but could give rise to the
same effect, independent of the nature of the companions.

\begin{figure*}[htb!]

\begin{center}
\rotatebox{90}{\includegraphics[width=3.in]{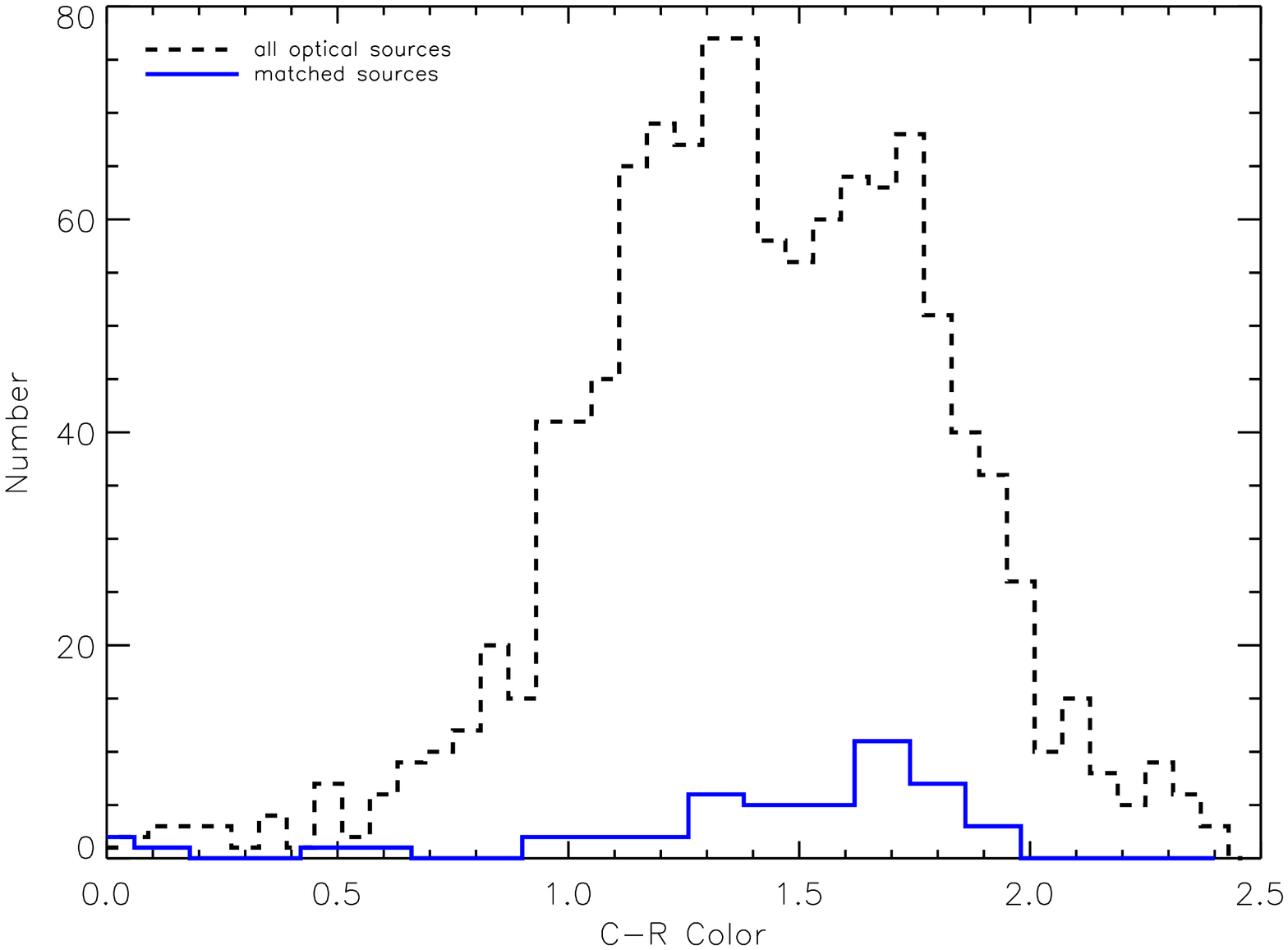}}
\rotatebox{90}{\includegraphics[width=3.in]{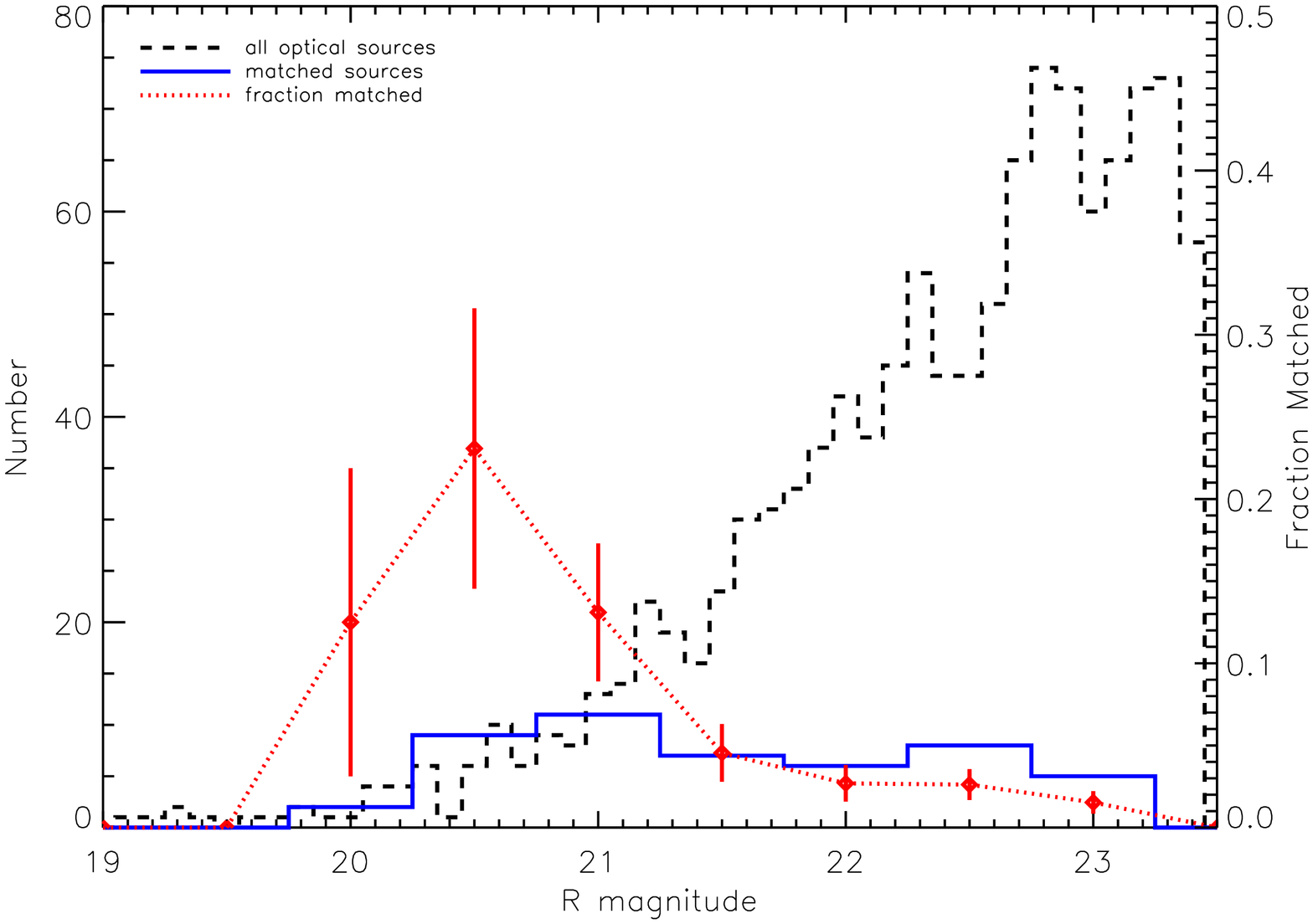}}
\end{center}

\caption{(Top:)
A histogram of the $C\!-\!R$ colors
of all globular cluster candidates (discrete
optical sources from Dirsch et al. (2005) with $R<23.5$ and $0 \leq
C-R \leq 2.5$),
found within $1.5^\prime \!<\! r \!<\! 6.^\prime$ of the center of
NGC~4636 (dashed histogram), plotted along with those that match with X-ray point sources
detected in the \cha\ observations (solid histogram). It is clear that
the X-ray point sources are preferentially associated with the redder
globular clusters. (Bottom:) Histograms for the same two samples 
(dashed-- all candidates, solid- those that match point sources)
as a function of GC magnitude. The red dotted line is the number of
matched GCs divided by the total number of GC candidates in each
luminosity bin.  This fraction declines towards fainter GC magnitude.}
\label{f:gcmatch}
\end{figure*}

We matched our list of X-ray point sources with the list of globular
cluster candidates, from the study of Dirsch et al. (2005), which uses
a deep mosaic CCD observation using the CTIO Blanco telescope.  The
photometry is available in the Washington $C$ and Kron-Cousins $R$
system.  Of the optical point sources listed in this work which extend
over an area of 0.25~deg$^2$ around the galaxy, we chose the sources
with magnitude $R<23.5$ and color $0\!<\! C-R\!<\! 2.5$ as globular
cluster candidates.  These are not spectroscopically confirmed to be
globular clusters belonging to the galaxy.  Figure~\ref{f:skymatch}
shows the distribution of both X-ray and optical point sources,
together with other background sources that are known in the field.
In a more recent paper, the group responsible for the original list
has published spectral observations of a small subset of 200 of the
original list of GC candidates (Schuberth et al. 2006). Of the sources
with magnitude $R<23.5$ and color $0<C-R<2.5$, $>$ 80\% of the
candidates were found to be globular clusters from measured redshift
and the rest foreground stars.

The positions of both X-ray and optical sources are known to an
accuracy of better than an arcsecond. We matched the two lists by
taking each X-ray point source, and finding its offset from the
nearest globular cluster candidate.  We checked for systematic
translation and rotation between the two lists by seeking to maximize
the matches for small values of rotation and translation of all
sources, but could not improve upon the matching done above.   Based on the distribution of offsets, we chose to limit the
search radius to 1.5$^{\prime\prime}$. If two GC candidates fell
within this radius, we assigned the X-ray source to the nearer one in
angular distance. Of the 318 sources in our list, 77 were matched to
globular cluster candidates. These sources are marked ``GC'' in the
last column of Table \ref{t:srclist} and are listed separately in
Table \ref{t:gcmatchlst}.  Of these 77 matched point sources, 48 lie
within $1.5^\prime \!<\! r \!<\! 6.0^\prime$ of the center of NGC~4636
and have more than 10 net counts in either of the ACIS-S or
combined ACIS-I observations.  Based on the number of X-ray
sources and GC candidates in this annulus, we expect only three chance
coincidences.  No other known background sources from NED
were matched with X-ray sources within 10$^{\prime}$ of the center of the
galaxy.
 
A $C\!-\!R$ color histogram of globular cluster candidates is shown in
the top panel of Figure~\ref{f:gcmatch} (dashed histogram).  As in many other early-type galaxies, the
distribution of the colors of candidate globular clusters here is bimodel.  Dirsch et al. (2005) fit a two-Gaussian model to their color
distribution, and find that the two peaks are at $C\!-\! R=1.28$ and
$1.77$, with the intervening minimum occurring at $C\!-\! R=1.5$. We
adopt these values, and here refer to those with $C\!-\! R>1.5$
as the ``redder'' GCs, and the rest, the ``bluer'' ones. Of the 48 matched
GCs between $1.5^\prime$ -- $6.0^\prime$ of the galactic center, 13
have colors between $0.85\! \le \! C\!-\! R \! < \! 1.5$, and 30 have
colors $ C\!-\! R \!\ge\! 1.5$, so 70\% of the matched GCs in Figure~\ref{f:gcmatch} with colors between 0.8 and 2.5 are in the
redder category.

The difference in color between the two populations is predominantly
due to a difference in metal abundance, since for stellar populations
that are more than a few Gyr old, the optical colors would be more
sensitive to the [Fe/H] index than to age (e.g., Worthey 1994, Bruzual
\& Charlot 1993).  In NGC~4636, even though bluer GCs are more
abundant, a majority of the X-ray point sources (LMXBs) are associated
with the redder GCs (those of near-solar abundance), as is shown by
the solid histogram in the left panel of  Figure~\ref{f:gcmatch}, which represents the color distribution of the
GCs matched with X-ray point sources.  A similar association
has been reported for other nearby early-type galaxies
(Kundu et al. 2002, Jord\'{a}n et al.2004,  
Kim et al. 2005, Kundu et al. 2007, Sivakoff et al. 2007).
This is also consistent with the
observation that most LMXBs associated with GCs in the Galaxy and M31 lie in those systems with a near-solar abundance (Grindlay 1993,
Bellazzini et al. 1995, Bregman 2006).

Other studies have found that X-ray point sources are preferentially
found in optically more luminous globular clusters (Angelini et
al. 2001, Sarazin et al. 2003,
Kundu et al. 2002, Jord\'{a}n et al.2004,  
Kim et al. 2005,  Xu et al. 2005). GCs containing LMXBs also have been noted
to be significantly denser in the Galaxy and in M31 (e.g. Bellazzini
et al. 1995). Since the mean size of a GC does not significantly vary
with luminosity (McLaughlin 2000), more luminous GCs can be expected
to be denser, and such a correlation can result
from the higher density
of potential companion stars in such systems. In the lower panel of Figure~\ref{f:gcmatch}, we show the distribution of all
GC candidates
$1.5^\prime \!<\! r \!<\! 6^\prime$ from the center of
NGC~4636 (dashed histogram), plotted along with those that match with X-ray point sources
(solid histogram).  The dotted line shows the ratio of matched to
total GC candidates as a function of R magnitude.  This ratio declines
towards fainter GCs, indicating that the probability of a GC containing an
LMXB is proportional to its luminosity, consistent with
other studies (e.g. Sarazin et al.2003, Jord{\'a}n et al.\ 2004, Sivakoff et al. 2007,
Woodley et al. 2008).

\begin{figure*}[htb!]\center\rotatebox{90}
{\includegraphics[width=4.0in]{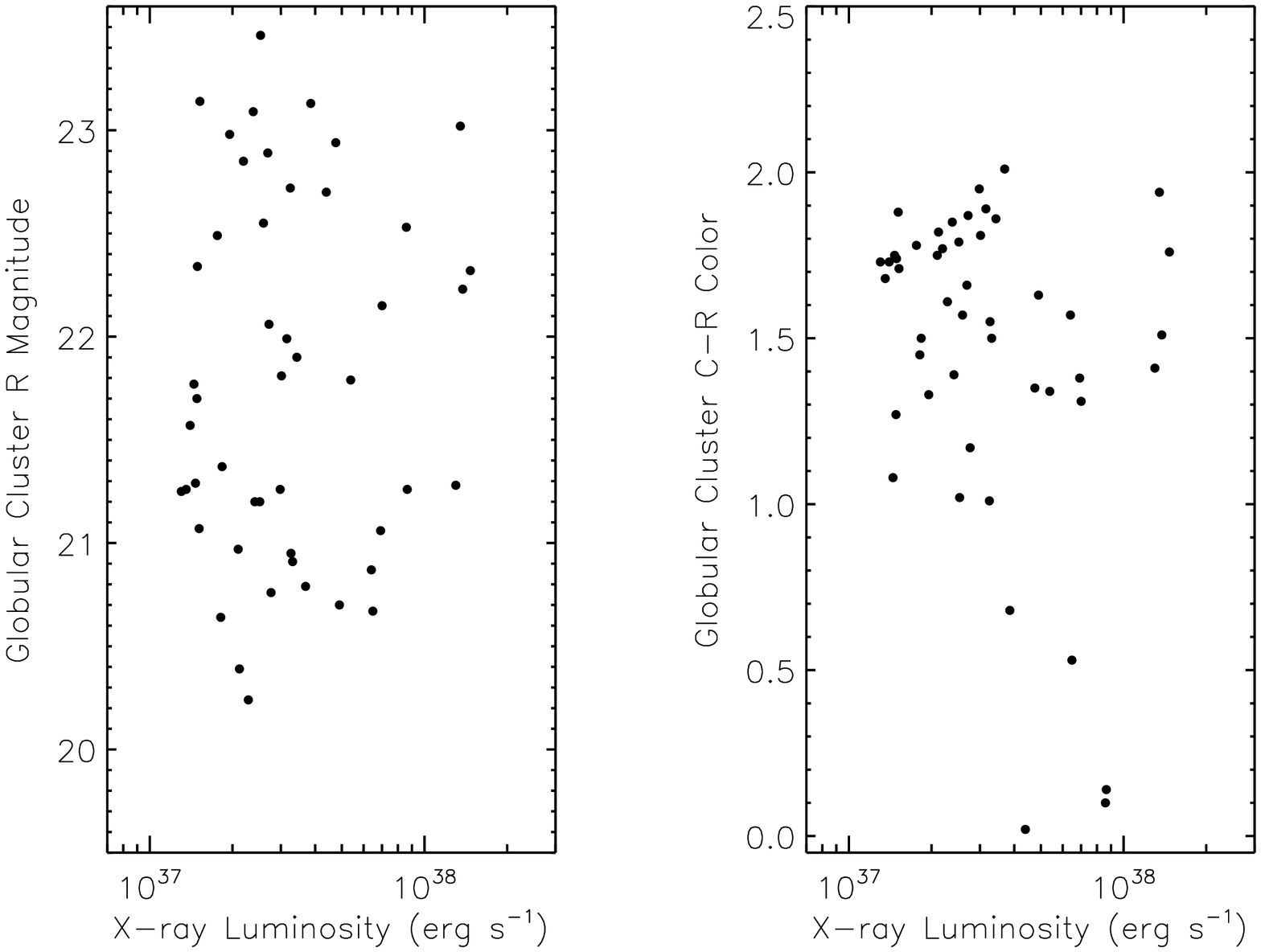}}
\caption{The mean X-ray luminosity of the point sources
with $\ge 10$ net counts that are found to match globular cluster candidates within 1.5$^\prime$ -- 6$^\prime$ of the galactic
center, plotted
against the $R$ magnitude (left) and the $C-R$ color (right) of the
matched globular cluster candidate. No trend is
observed with the magnitude of the host  globular cluster, while, as
seen in  Figure~\ref{f:gcmatch}, the overabundance of matched X-ray
sources in the redder GCs is easily apparent.
\label{f:lxt1mag}}
\end{figure*}

In Figure~\ref{f:lxt1mag}, we plot the apparent magnitude and color of
the host globular clusters against the X-ray luminosity of the matched
X-ray point sources. These plots indicate that
the X-ray
luminosity of the matched point sources 
does not depend on the color or absolute luminosity
of the host GCs, which agrees with earlier work noted above. 
Together with the previous observation, this should be a useful
constraint on the formation process of LMXBs in globular clusters.

\begin{figure*}[htb!]
\begin{center}
\rotatebox{90}{\includegraphics[width=3.5in]{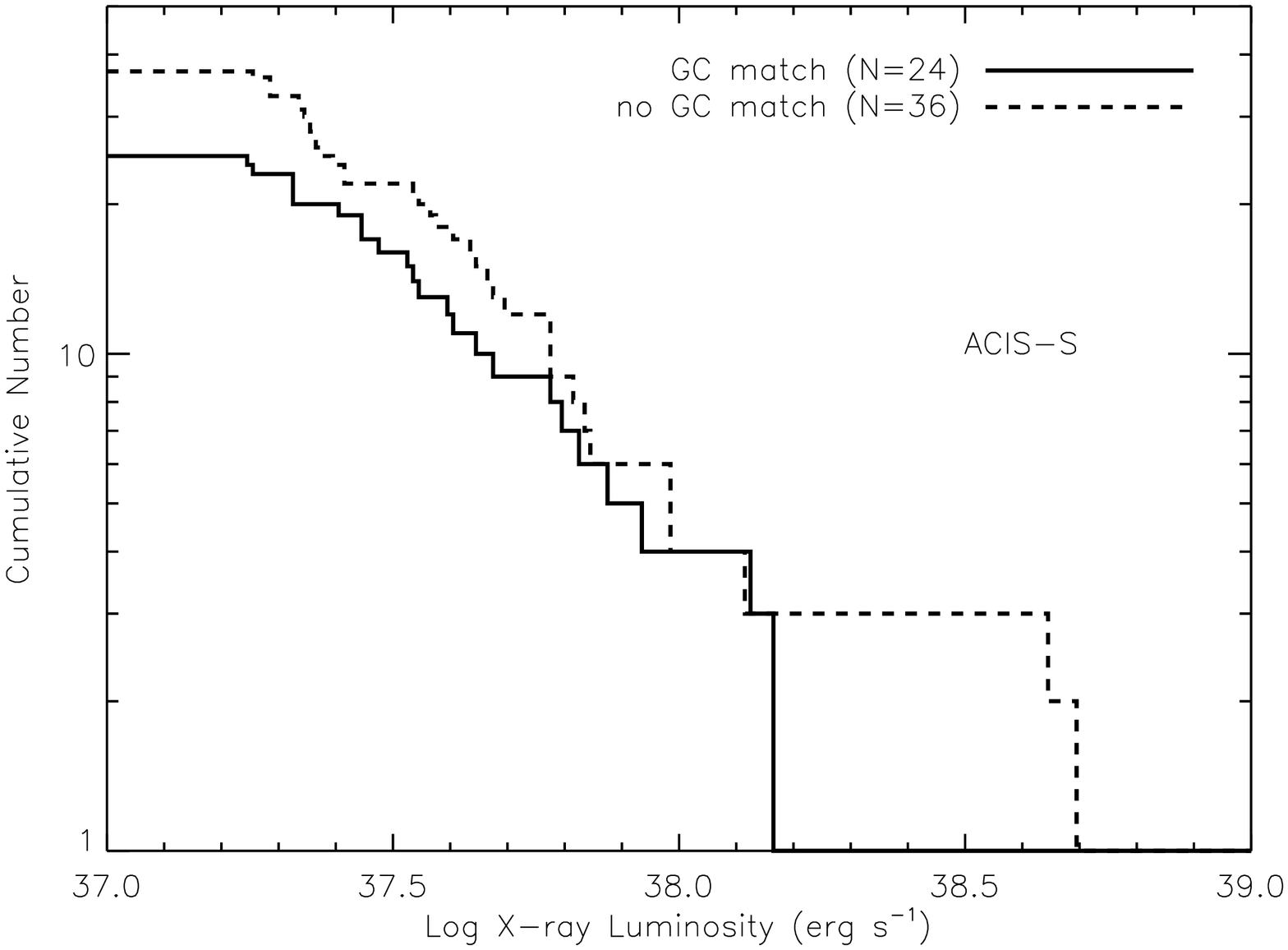}}
\rotatebox{90}{\includegraphics[width=3.5in]{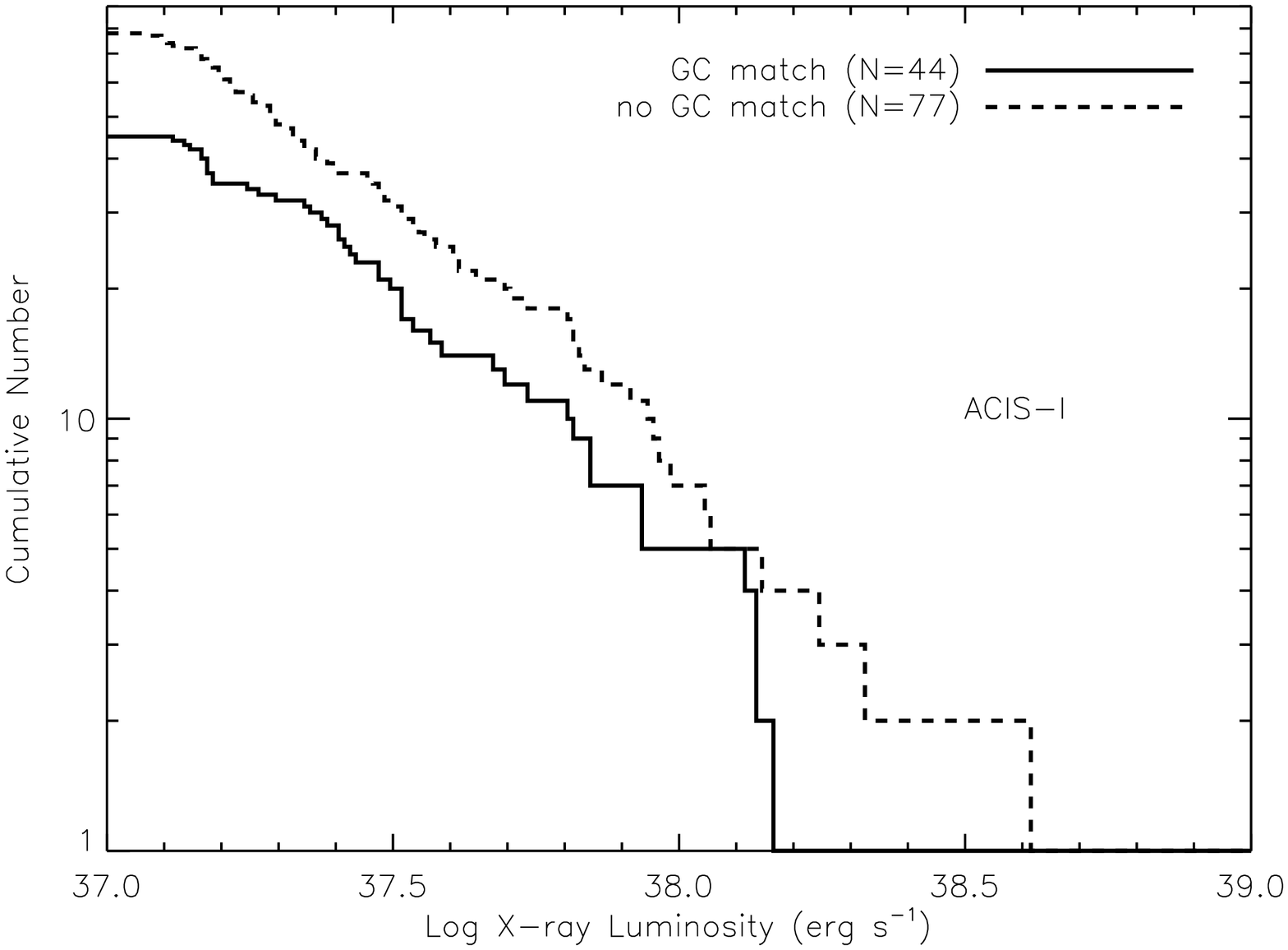}}
\end{center}
\caption{The luminosity functions (LF) of all X-ray point sources,
found within $1.5^\prime \!<\! r \!<\! 6.0^\prime$ of the center of
NGC 4636 with $\geq$ 10 net counts that are matched with globular
cluster candidates (solid histogram). Also plotted are the LFs of all X-ray point sources that did not match a GC candidate
(dashed histogram). The two plots are for 0.5-2 keV band luminosities calculated for
the ACIS-S Sequence 600083 observation (top) and the combined 
ACIS-I Sequence 600300 and
600331 observations (bottom).  In both cases the LF slopes for the
matched and unmatched sources are similar between $1.8\times10^{37}$ erg s$^{-1}
\leq L_{x} \leq 1\times10^{38}$ erg s$^{-1}$ (log=37.25-38).  Note
that the brightest X-ray sources are not matched with globular clusters.}
\label{f:lx-inout}
\end{figure*}

Finally, we compare the luminosity function (LF) of the X-ray point sources
matched with globular cluster candidates to the LF of those that are not
matched with any optical source. Here, we study only the point sources
within $1.5^\prime \!<\! r \!<\! 6.0^\prime$ of the center of
NGC~4636, where the X-ray sources are not seriously contaminated by
background AGN (see Figure \ref{f:surfden}).  In
Figure~\ref{f:lx-inout}, 
we plot the LFs of GC-matched and non-GC
matched X-ray point sources. The two plots represent X-ray luminosities obtained
from two different epochs of observations,  to account for the
possible change in LF due to long term variability in these sources.
The solid histogram shows the LF of the matched point sources, and the
dashed histogram is the LF of all X-ray point sources that did not
match a GC candidate. Note that the brightest X-ray sources
are not matched with any GC candidate.  In both cases, the slopes of the LFs of the GC sources
and non-GC sources in the range $1.8\times10^{37}$ erg s$^{-1}
\leq L_{x} \leq 1\times10^{38}$ erg s$^{-1}$ match within 2$\sigma$.  For the ACIS-S observation, from a Kolmogorov-Smirnov (K-S) test,
the probability that the two samples are drawn from the same
population is 78\%, whereas for the ACIS-I samples, the corresponding
value is 50\%, with the maximum deviation occurring at the faint end
in both cases.  We note that the K-S test underestimates this probability
for discrete distributions (see, for example, Sheskin 2003). At the faint end of the luminosity
function, there is emerging evidence, from nearby galaxies
like Cen A (NGC~5128) where more complete samples of fainter sources can be studied, 
that there are fewer X-ray faint sources found in GCs than X-ray bright
ones (e.g. Woodley et al. 2008).  We see perhaps some evidence for a
dearth of GC matches at lower luminosities in the ACIS-I LFs (Figure
~\ref{f:lx-inout}, bottom panel).

\begin{figure*}[htb!]\center\rotatebox{90}
{\includegraphics[width=4.5in]{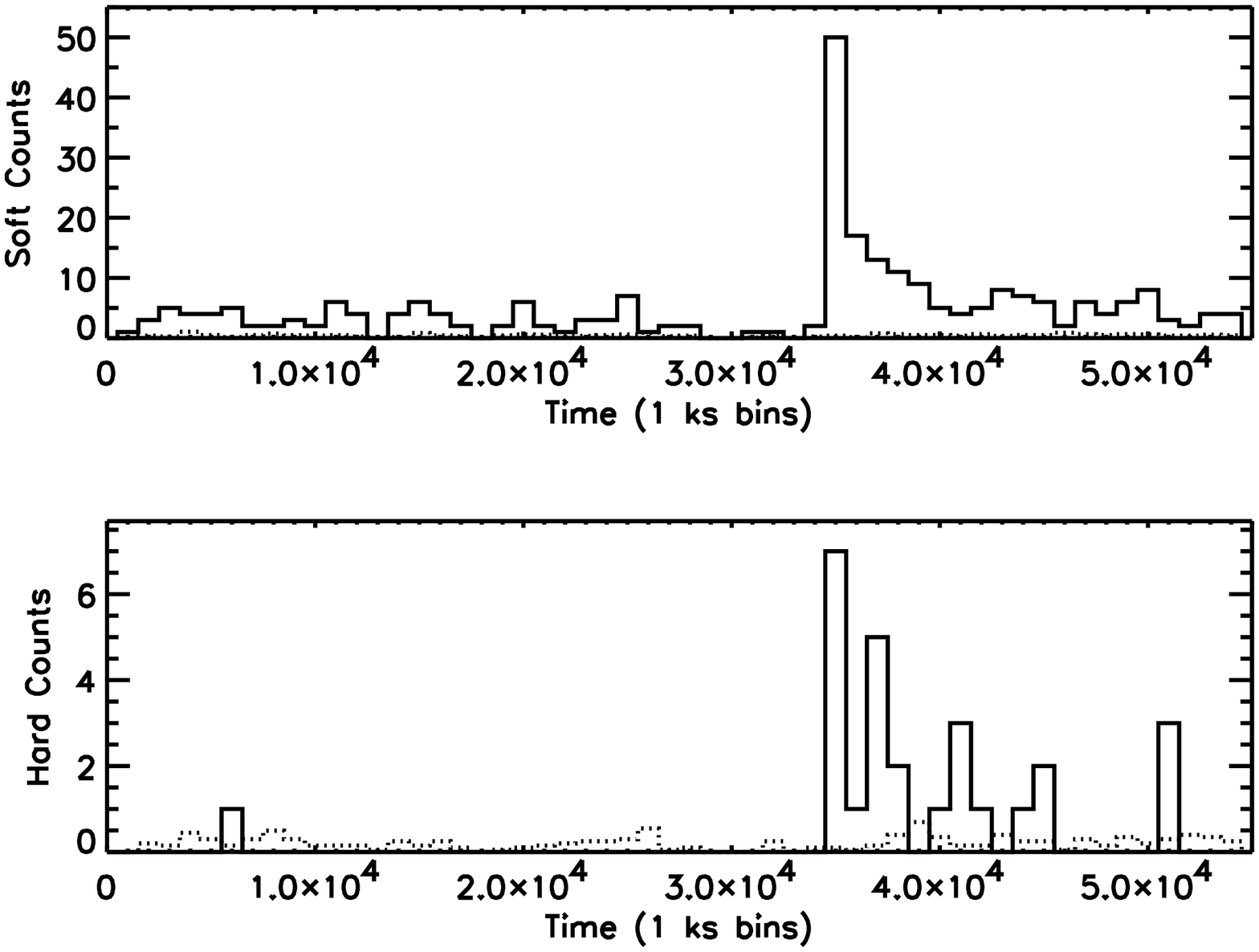}}
\caption{Soft band (0.3 -- 2 keV, top panel) and hard band (2 -- 6
  keV, bottom panel) lightcurves from the ACIS-S observation for the
  bursting X-ray source at $\alpha(J2000) = 190.8825^{\circ}$,
  $\delta(J2000) = 2.6190^{\circ}$.  Background counts are shown by
  the dotted line in each panel.
\label{f:var1_ratio}}
\end{figure*}

\subsection{Individual Sources}
\label{s:individ}

\subsubsection{Galactic Nucleus}
\label{s:nucleus}
We detect a soft, luminous ($L_{X} = 2.2\times10^{38}$ erg s$^{-1}$)
X-ray point source at the galaxy's nucleus, as determined from the VLA
5~GHz (Griffith et al. 1995)
radio position ($\alpha$(J2000) = 190.7077, $\delta$(J2000) =
2.6878).  Loewenstein et al. (2001) do not detect nuclear activity in NGC 4636 in
the 2 -- 10 keV band.  However, this source is very soft, and
$\sim90\%$ of its emission falls below the 2 -- 10 keV band.  
This source is shown with a purple X in the color-color diagram
(Figure \ref{f:colcol}) and is significantly softer than the LMXBs.
The spectrum is well-fit by a power-law model with a photon index of
$\lambda$ = 2.36 (68\% confidence range is 1.94 -- 2.76).  In fellow
Virgo elliptical galaxies NGC 4472 and NGC 4649, Soldatenkov et
al. (2003) detect two soft x-ray sources at the galactic nuclei with similar power
law indices (2.5 $\pm$ 0.4 and $<$ 2.2).

\subsubsection{Additional Black Hole Candidates}
\label{s:blackhole}

We detected two soft, luminous ($L_{x} \approx 2\times10^{38}$ erg
s$^{-1}$) sources very near the center of the galaxy ($d < 5^{\prime\prime}$), whose soft spectra, like the source at the nucleus,
may be signatures of black holes.  These
sources are shown by the purple asterisk and plus sign in the color-color diagram (Figure
\ref{f:colcol}).  McClintock \& Remillard (2004) find
that black hole binaries in the thermal-dominated state are
well-described by a disk-blackbody model with T = 1 keV.  The spectra of the two sources are very similar,
so we added them before fitting with XSPEC (Arnaud 1996).  With a disk-blackbody
model, we find a best-fit temperature of 0.30 keV with a 68\% confidence range of 0.27 -- 0.33 keV.  The co-added spectra are
equally well-fit by a power-law model with a photon index of $\lambda$
= 2.81 (68\% confidence range 2.58 -- 3.03).  

Although we are prevented by the insufficient source counts from
reaching a definitive conclusion, it is possible that these sources,
and the one at the nucleus, are indeed signatures of black holes at
the center of NGC 4636.  If so, they may be massive black holes which
fell to the galaxy center under dynamical friction, in which case they
would have masses greater than $10^{5}$ M$_{\odot}$ (Tremaine \&
Ostriker 1999).  However, given the luminosities of these sources, it
is more likely that they are remnants of merged galaxies, similar to
what Komossa et al. (2003) observe in the elliptical galaxy NGC 6240.

\subsubsection{SN 1939A}
\label{s:sn1939a}

SN 1939A is a Type 1a supernova 0.57$^{\prime}$ from the center of NGC 4636
($\alpha$(J2000)=190.700$^{\circ}$, $\delta$(J2000)=2.693$^{\circ}$
(Zwicky 1939, Giclas 1939, Tsvetkov \& Bartunov
1993).  We do not detect X-ray
emission at this location.  By generating a large number of Poisson
realizations from the observed background at this location in our
\cha\ observations, we calculate a 3$\sigma$ X-ray count rate upper
limit of 2.5$\times$10$^{-4}$ cts s$^{-1}$ in the 0.5 -- 2 keV band.
This translates to a luminosity upper limit of $\sim 4.5\times10^{37}$
erg s$^{-1}$ in the same energy band.

\subsubsection{Bursting X-ray Source}
\label{s:burst}

A light curve for the bursting X-ray source, detected at
$\alpha(J2000) = 190.8825^{\circ}$, $\delta(J2000) = 2.6190^{\circ}$,
is shown in Figure \ref{f:var1}.  The matching source (0.8$^{\prime\prime}$ away) in the 2MASS
All-Sky Catalog of Point Sources, 2MASS J12433175+0237080
($\alpha(J2000) = 190.882303^{\circ}$, $\delta(J2000) =
2.618898^{\circ}$) has J, H, and K band magnitudes of 10.139 $\pm$
0.032, 9.485 $\pm$ 0.03, and 9.293 $\pm$ 0.028, respectively (Skrutskie 2006).
From the Sloan Digital Sky Survey Data Release 6, we get g and r band
magnitudes of 13.92 $\pm$ 0.01 and 14.83 $\pm$ 0.01, respectively, for matching source SDSS J124331.75+023708.0
($\alpha(J2000) = 190.8823016^{\circ}$, $\delta(J2000) =
2.61889293^{\circ}$, separation=0.8$^{\prime\prime}$) (Adelman-McCarthy 2008).
Based on the source's optical and infrared brightness and its large
distance from the center of the galaxy (11.3$^\prime$), it is most
likely not a member of NGC 4636, but is probably a relatively
nearby flare star in our Galaxy.

Figure \ref{f:var1_ratio} shows hard and soft band lightcurves for
this source from the ACIS-S observation.  No hard (2 -- 6 keV) counts
are detected above the background prior to the burst.

\section{Summary and Conclusions}
\label{s:summary}

We have analyzed three \cha\ ACIS observations, taken over three years
and totaling $\sim$193 ks, of the Virgo galaxy NGC 4636, and
have detected 277 X-ray point sources above a luminosity of
$1.2\times10^{37}$ erg s$^{-1}$ in the 0.5 -- 2 keV band, outside the
central $1.5^{\prime}$ bright galaxy core.  Based on the estimated 
density of AGN in the field, $\sim$123 of these are likely members of the
galaxy, while the rest are likely AGN.  In the region from
1.5$^\prime$ to 6$^\prime$ from the center of the galaxy, there are
129 X-ray point sources detected with $\geq$ 10 net counts, $\sim$25\% of
which are likely AGN.

We calculate X-ray colors from fluxes (rather than from counts, as is
traditionally done), and find that this results in more clearly
grouped populations on the X-ray color-color diagram.  We identify a
large group of LMXBs, and a small group of much softer sources,
including three within 5$^{\prime\prime}$ of the galactic center which
may be black holes.

We find 77 matches between our X-ray point sources and potential
globular cluster (GC) candidates found in deep optical images of
NGC~4636. Choosing the subset of 48 matched point sources with $\ge$10
net counts that correlate with GC candidates and lie within $1.5^\prime \!<\! r \!<\! 6^\prime$ of the
center of the galaxy (out of 129 X-ray point sources in this annulus),
we find that the overwhelming majority are associated with the redder
GC candidates, those that are thought to have near-solar metal
abundance.  We find that the fraction of GC candidates with X-ray
point source matches decreases with decreasing GC luminosity.
The luminosity functions of the point sources matched with
GCs and of those that are unmatched have similar slopes.
We searched for variable sources on timescales ranging from hours to
years.  We find that 54 sources in the common field of view (24\%)
show long-term variability and vary by $\geq 3\sigma$ between the
January 2000 ACIS-S observation and the February 2003 ACIS-I
observations, while about 10\% of sources show short-term variability
within a single observation.

\acknowledgments

We thank Vinay Kashyap for helpful discussions on short term source
variability and the development of his extremely useful
\textit{timevarvk} IDL code.  This work was
supported by NASA contracts NAS8-39073 and NAS8-03060, the Chandra
Science Center, the Smithsonian Institution, and the University of
Birmingham.

This research has made use of SAOImage DS9, developed by Smithsonian Astrophysical Observatory.

This research has made use of the VizieR catalogue access tool, CDS,
Strasbourg, France. VizieR is a joint effort of CDS (Centre de Données
astronomiques de Strasbourg) and ESA-ESRIN (Information Systems
Division).




\clearpage
\pagebreak
\LongTables 
\begin{deluxetable}{llcccccl}
\tablecolumns{8}
\tabletypesize{\footnotesize}
\tablecaption{Summary of X-ray point sources detected in NGC 4636 \label{t:srclist}} 
\tablewidth{0pt}
\tablehead{
\colhead{$\alpha$(J2000)} & \colhead{$\delta$(J2000)} & \colhead{Net
  Cts\tablenotemark{a}} & \colhead{Luminosity\tablenotemark{a}} & \colhead{Soft Color\tablenotemark{b}} &
\colhead{Hard Color\tablenotemark{c}} & \colhead{Dist. from Center ($^{\prime}$)} & \colhead{Notes\tablenotemark{d}} 
}
\startdata
190.7077 & 2.6878 & 202 & 2.202E+38 & -0.39 & -0.05 &  0.00 & 3  MV\\
190.7071 & 2.6867 & 138 & 2.375E+38 & -0.67 & -0.09 &  0.08 & 3  V,MV\\
190.7091 & 2.6876 & 157 & 1.718E+38 & -0.47 & -0.08 &  0.08 & 3    \\
190.7066 & 2.6918 &  68 & 7.432E+37 & -0.69 & -0.11 &  0.25 & 2  V\\
190.7016 & 2.6860 &  45 & 5.044E+37 &  0.24 &  0.28 &  0.38 & 2    \\
190.7068 & 2.6811 &  56 & 9.761E+37 & -0.19 & -0.31 &  0.40 & 1  V\\
190.7158 & 2.6862 &  44 & 4.774E+37 &  0.17 &  0.60 &  0.49 & 3    \\
190.7047 & 2.6796 &  43 & 4.730E+37 & -0.23 & -0.02 &  0.52 & 3  GC 19\\
190.7134 & 2.6774 &  45 & 5.000E+37 &  0.09 &  0.47 &  0.71 & 2  V,GC 77\\
190.7195 & 2.6854 &  22 & 3.948E+37 & -0.02 &  0.33 &  0.72 & 1  GC 24\\
190.7198 & 2.6884 &  53 & 5.807E+37 &  0.07 &  0.39 &  0.72 & 2  V\\
190.7180 & 2.6947 &  56 & 6.047E+37 &  0.10 &  0.41 &  0.74 & 3    \\
190.7150 & 2.6772 &  38 & 4.210E+37 &  0.02 &  0.38 &  0.77 & 3    \\
190.7152 & 2.6756 &  29 & 5.147E+37 &  0.06 &  0.40 &  0.86 & 3    \\
190.6956 & 2.6975 &  48 & 5.367E+37 &  0.07 &  0.13 &  0.93 & 2  V\\
190.7091 & 2.7035 &  36 & 3.926E+37 &  0.12 &  0.42 &  0.95 & 2    \\
190.7254 & 2.6887 &  21 & 3.663E+37 &  0.02 &  0.63 &  1.06 & 1    \\
190.7262 & 2.6892 &  34 & 3.676E+37 & -0.20 &  0.38 &  1.11 & 2    \\
190.7250 & 2.6979 &  21 & 2.253E+37 &  0.02 &  0.90 &  1.20 & 2  GC 46\\
190.6877 & 2.6931 &  19 & 2.228E+37 &  0.45 & -0.37 &  1.24 & 2  GC 62\\
190.7102 & 2.6670 &  29 & 3.239E+37 & -0.11 &  0.07 &  1.26 & 2  V\\
190.6902 & 2.6993 &  42 & 4.782E+37 &  0.05 &  0.29 &  1.26 & 3  SV-I\\
190.6962 & 2.7055 &  27 & 4.679E+37 &  0.24 &  0.06 &  1.27 & 1  V   \\
190.7296 & 2.6897 &  71 & 7.588E+37 &  0.05 &  0.71 &  1.32 & 3  GC 26\\
190.7212 & 2.6696 &  13 & 1.487E+37 & -0.16 & -0.12 &  1.36 & 2    \\
190.7295 & 2.6804 &  17 & 1.840E+37 & -0.09 &  0.49 &  1.38 & 2    \\
190.6847 & 2.6889 &  57 & 6.500E+37 & -0.04 &  0.41 &  1.38 & 3    \\
190.7027 & 2.7108 &  19 & 2.171E+37 & -0.00 & -0.14 &  1.41 & 2    \\
190.6833 & 2.6826 &  36 & 4.080E+37 &  0.06 &  0.50 &  1.50 & 3    \\
190.7194 & 2.7107 &  36 & 3.853E+37 &  0.12 &  0.75 &  1.54 & 3  GC 21\\
190.7315 & 2.6753 &  15 & 1.690E+37 &  0.46 & -0.37 &  1.61 & 2    \\
190.7325 & 2.6766 &  90 & 9.686E+37 &  0.28 &  0.44 &  1.63 & 3  V   \\
190.7149 & 2.7141 &  15 & 1.686E+37 & -0.12 & -0.14 &  1.63 & 2    \\
190.7268 & 2.7077 &  13 & 1.447E+37 &  0.09 &  0.45 &  1.65 & 2  GC 65\\
190.6979 & 2.6619 & 186 & 2.101E+38 & -0.12 & -0.19 &  1.66 & 3  V   \\
190.7211 & 2.7123 &  27 & 2.983E+37 &  0.09 &  0.48 &  1.68 & 3  GC 13\\
190.7093 & 2.7161 &  67 & 7.291E+37 &  0.06 &  0.55 &  1.70 & 3    \\
190.7369 & 2.6863 &  60 & 6.404E+37 &  0.12 &  0.41 &  1.75 & 3
SV-S,GC 20\\
190.7333 & 2.7030 &  14 & 1.521E+37 & -0.05 &  0.25 &  1.78 & 2  GC 44\\
190.6787 & 2.6817 &  22 & 2.530E+37 &  0.03 &  0.57 &  1.78 & 3  GC 22\\
190.6787 & 2.6805 &  28 & 3.246E+37 & -0.08 &  0.53 &  1.80 & 3  GC 12\\
190.7367 & 2.6795 &  16 & 1.813E+37 &  0.08 &  0.82 &  1.81 & 2    \\
190.6783 & 2.6982 &  79 & 9.043E+37 &  0.13 &  0.75 &  1.88 & 3  V\\
190.7134 & 2.6567 &  20 & 3.640E+37 &  0.23 &  0.30 &  1.90 & 3  \\
190.6927 & 2.7157 &  12 & 1.356E+37 &  0.15 &  0.09 &  1.90 & 2  GC 54\\
190.6880 & 2.7132 &  59 & 6.648E+37 & -0.05 &  0.52 &  1.93 & 3    \\
190.7178 & 2.7188 &  11 & 1.252E+37 &  0.33 &  0.21 &  1.96 & 2    \\
190.6781 & 2.7048 &  12 & 2.119E+37 & -0.13 &  0.42 &  2.05 & 1
SV-S,GC 28\\
190.7007 & 2.7216 &  13 & 1.429E+37 & -0.25 &  0.24 &  2.07 & 2    \\
190.7433 & 2.6861 &  31 & 3.399E+37 &  0.07 &  0.47 &  2.13 & 3    \\
190.6947 & 2.7209 &  14 & 1.567E+37 &  0.01 &  0.16 &  2.13 & 2    \\
190.7095 & 2.7236 &  11 & 1.925E+37 & -0.02 &  0.52 &  2.15 & 1    \\
190.7226 & 2.6550 &  11 & 1.308E+37 &  0.06 &  0.17 &  2.16 & 2  \\
190.7126 & 2.7235 &  13 & 2.350E+37 &  0.05 &  0.41 &  2.16 & 1    \\
190.7398 & 2.7059 &  23 & 2.473E+37 &  0.11 &  0.73 &  2.21 & 2  V\\
190.6907 & 2.7206 &  13 & 1.512E+37 &  0.28 &  0.15 &  2.22 & 3  GC 37\\
190.6807 & 2.7141 &  19 & 2.213E+37 & -0.47 &  0.08 &  2.27 & 2  V\\
190.7340 & 2.7155 &  23 & 2.523E+37 &  0.05 & -0.15 &  2.29 & 2  V,MV,SV-I\\
190.7081 & 2.7260 &  14 & 1.523E+37 &  0.07 &  0.39 &  2.29 & 3    \\
190.6797 & 2.7138 &  10 & 1.812E+37 &  0.35 &  0.10 &  2.30 & 3  GC 31\\
190.6718 & 2.6728 & 150 & 1.753E+38 &  0.07 &  0.36 &  2.34 & 3    \\
190.7468 & 2.6811 &  27 & 2.927E+37 &  0.04 &  0.36 &  2.38 & 3    \\
190.6675 & 2.6818 &  38 & 6.769E+37 & -1.26 & -0.01 &  2.44 & 3    \\
190.7068 & 2.7288 &  34 & 3.690E+37 &  0.14 &  0.11 &  2.46 & 3  GC 14\\
190.6695 & 2.6725 &  94 & 1.105E+38 & -0.07 &  0.18 &  2.47 & 3    \\
190.7400 & 2.6611 &  34 & 3.772E+37 &  0.05 &  0.43 &  2.51 & 3  \\
190.7452 & 2.6685 &  45 & 4.901E+37 &  0.05 &  0.40 &  2.53 & 3  V\\
190.7021 & 2.7296 &  13 & 1.475E+37 &  0.15 &  0.08 &  2.53 & 2    \\
190.7496 & 2.6800 &  24 & 2.592E+37 &  0.26 & -0.30 &  2.55 & 2  V,GC 59\\
190.7145 & 2.6435 & 118 & 1.351E+38 &  0.06 &  0.44 &  2.69 & 3  GC 16\\
190.6774 & 2.7211 &  76 & 8.652E+37 &  0.04 &  0.31 &  2.70 & 2  V,GC 58\\
190.6779 & 2.7223 &  41 & 4.753E+37 &  0.04 &  0.58 &  2.74 & 3  GC 35\\
190.6965 & 2.7323 &  29 & 3.292E+37 & -0.06 &  0.56 &  2.75 & 2  V   \\
190.6685 & 2.7120 &  36 & 4.167E+37 &  0.09 &  0.45 &  2.77 & 2  V   \\
190.6644 & 2.7046 &  23 & 2.715E+37 & -0.00 &  0.49 &  2.79 & 2  GC 53\\
190.7320 & 2.6479 &  15 & 2.762E+37 & -0.05 &  0.67 &  2.80 & 3  GC 15\\
190.7264 & 2.6437 &  14 & 1.627E+37 & -0.10 &  0.32 &  2.88 & 3    \\
190.7549 & 2.7000 &  25 & 4.458E+37 &  0.15 &  0.53 &  2.92 & 1  V   \\
190.6601 & 2.6978 &  27 & 3.265E+37 &  0.04 &  0.35 &  2.92 & 3  GC 8\\
190.7556 & 2.6970 &  50 & 5.386E+37 &  0.06 &  0.49 &  2.92 & 3  GC 36\\
190.7575 & 2.6918 &  12 & 2.180E+37 &  0.08 & -0.22 &  3.00 & 1  V\\
190.7348 & 2.6449 &  11 & 2.097E+37 & -0.04 &  0.86 &  3.04 & 3  GC 32\\
190.7499 & 2.6594 & 247 & 4.455E+38 &  0.12 &  0.58 &  3.05 & 3  V   \\
190.6569 & 2.6927 &  26 & 3.153E+37 &  0.03 &  0.36 &  3.06 & 3  GC 5\\
190.7209 & 2.6372 &  98 & 1.127E+38 & -0.58 & -0.13 &  3.14 & 3    \\
190.6809 & 2.6417 &  57 & 6.921E+37 &  0.02 &  0.33 &  3.20 & 3  GC 25\\
190.6740 & 2.6463 &  77 & 9.297E+37 &  0.04 &  0.45 &  3.21 & 3  V   \\
190.6699 & 2.6470 &  12 & 1.484E+37 &  0.25 &  0.59 &  3.34 & 2  GC 61\\
190.7042 & 2.6311 &  14 & 2.627E+37 &  0.11 &  0.30 &  3.41 & 3  \\
190.7647 & 2.6900 &  60 & 6.485E+37 &  0.15 &  0.40 &  3.42 & 3  GC 6\\
190.7647 & 2.6847 &  20 & 2.191E+37 &  0.21 &  0.41 &  3.42 & 3  GC 17\\
190.7618 & 2.6691 &  11 & 1.280E+37 &  0.37 &  0.10 &  3.43 & 2  MV\\
190.7055 & 2.7454 &  10 & 1.184E+37 &  0.32 &  0.27 &  3.46 & 2    \\
190.6963 & 2.7446 &  21 & 2.329E+37 & -0.86 &  0.04 &  3.47 & 3    \\
190.7206 & 2.6306 & 358 & 4.167E+38 & -0.24 & -0.18 &  3.52 & 3    \\
190.7664 & 2.6852 &  16 & 1.797E+37 &  0.06 &  0.51 &  3.52 & 2  \\
190.7227 & 2.7458 &  61 & 6.500E+37 &  0.02 & -0.14 &  3.59 & 2    \\
190.7646 & 2.6689 &  24 & 2.688E+37 &  0.35 &  0.27 &  3.59 & 3  GC 34\\
190.7688 & 2.7011 &  28 & 3.068E+37 &  0.08 &  0.75 &  3.75 & 3  SV-I\\
190.7676 & 2.6672 &  14 & 1.566E+37 &  0.06 &  0.49 &  3.80 & 3    \\
190.7712 & 2.6908 &  12 & 1.302E+37 &  0.12 &  0.61 &  3.81 & 3  GC 3\\
190.7715 & 2.6879 &  33 & 3.538E+37 &  0.05 &  0.30 &  3.83 & 3    \\
190.6451 & 2.6996 &  12 & 1.556E+37 &  0.03 &  0.48 &  3.83 & 2  \\
190.7625 & 2.6541 &  63 & 7.007E+37 &  0.04 &  0.51 &  3.86 & 3  GC 7\\
190.7428 & 2.6333 &  17 & 1.960E+37 &  0.09 &  0.74 &  3.89 & 3  SV-I\\
190.6608 & 2.7329 &  18 & 2.131E+37 & -0.21 &  0.04 &  3.91 & 2    \\
190.7773 & 2.6984 &  16 & 1.761E+37 &  0.03 &  0.51 &  4.22 & 2  GC 51\\
190.6769 & 2.7526 & 120 & 1.378E+38 &  0.17 &  0.31 &  4.30 & 3  SV-I,GC 2\\
190.7552 & 2.7421 &  83 & 8.723E+37 &  0.16 &  0.29 &  4.32 & 3    \\
190.7796 & 2.6748 &  16 & 1.833E+37 &  0.09 &  0.68 &  4.38 & 2  GC 66\\
190.7128 & 2.7613 &  12 & 1.323E+37 &  0.41 & -0.22 &  4.42 & 2    \\
190.7231 & 2.6155 &  12 & 1.466E+37 & -0.12 &  0.86 &  4.43 & 2  GC 42\\
190.6344 & 2.6966 &  40 & 5.025E+37 &  0.08 &  0.62 &  4.43 & 2    \\
190.7825 & 2.6849 &  17 & 1.905E+37 & -0.28 &  0.09 &  4.49 & 2    \\
190.6790 & 2.7570 &  12 & 1.413E+37 &  0.06 &  0.78 &  4.50 & 2    \\
190.6494 & 2.7349 &  26 & 3.193E+37 &  0.01 &  0.10 &  4.50 & 2    \\
190.6447 & 2.7308 &  19 & 2.411E+37 &  0.19 &  0.12 &  4.58 & 2  GC 40\\
190.7469 & 2.7535 &  19 & 2.081E+37 &  0.13 &  0.39 &  4.59 & 2    \\
190.6821 & 2.7600 &  57 & 6.509E+37 & -0.18 & -0.21 &  4.60 & 2  V\\
190.7004 & 2.7646 &  17 & 1.952E+37 &  0.17 &  0.21 &  4.63 & 3  GC 1\\
190.7056 & 2.7652 & 117 & 1.300E+38 &  0.06 &  0.54 &  4.65 & 3  GC 29\\
190.7309 & 2.6129 &  45 & 5.396E+37 &  0.02 &  0.52 &  4.70 & 3    \\
190.6667 & 2.7547 &  20 & 2.426E+37 & -0.45 & -0.04 &  4.71 & 2    \\
190.6925 & 2.7648 &  29 & 3.309E+37 &  0.13 &  0.58 &  4.71 & 3  GC 11\\
190.6433 & 2.7329 &  12 & 1.519E+37 & -0.05 &  0.23 &  4.72 & 2    \\
190.6289 & 2.6775 &  23 & 3.015E+37 & -0.03 &  0.43 &  4.77 & 2  GC 41\\
190.7080 & 2.6068 &  13 & 2.513E+37 &  0.01 &  0.60 &  4.86 & 3  SV-S,GC 30\\
190.7533 & 2.6181 &  12 & 1.509E+37 &  0.05 &  0.25 &  5.00 & 2    \\
190.6708 & 2.6102 &  23 & 4.389E+37 &  0.04 &  0.51 &  5.16 & 3  GC 33\\
190.6772 & 2.7682 &  24 & 2.829E+37 &  0.05 &  0.40 &  5.16 & 3    \\
190.6942 & 2.7729 &  12 & 1.402E+37 &  0.20 & -0.11 &  5.17 & 2  GC 76\\
190.6229 & 2.7059 &  66 & 8.587E+37 &  0.05 &  0.42 &  5.20 & 2
MV,SV-I,GC 69\\
190.6739 & 2.7687 &  18 & 2.169E+37 & -0.76 & -0.14 &  5.26 & 2    \\
190.6947 & 2.7748 &  19 & 2.194E+37 &  0.24 &  0.65 &  5.28 & 2    \\
190.7299 & 2.7745 &  31 & 3.429E+37 &  0.06 &  0.30 &  5.37 & 2  V,GC 68\\
190.6633 & 2.6097 &  13 & 1.776E+37 &  0.02 &  0.67 &  5.39 & 3    \\
190.7489 & 2.6077 &  10 & 1.281E+37 &  0.22 &  0.07 &  5.40 & 2    \\
190.7923 & 2.7210 & 126 & 1.381E+38 &  0.22 &  0.70 &  5.45 & 2    \\
190.7172 & 2.7783 &  57 & 6.405E+37 &  0.11 &  0.43 &  5.46 & 3  SV-S\\
190.7938 & 2.7207 & 134 & 1.468E+38 &  0.12 &  0.42 &  5.53 & 2  GC 56\\
190.6608 & 2.7673 &  17 & 2.131E+37 &  0.15 &  0.77 &  5.54 & 2    \\
190.7187 & 2.7797 &  26 & 2.969E+37 & -0.13 &  0.24 &  5.55 & 2    \\
190.6781 & 2.7758 &  59 & 6.904E+37 &  0.11 &  0.39 &  5.57 & 3    \\
190.7851 & 2.6359 &  20 & 2.379E+37 &  0.05 &  0.19 &  5.59 & 2  GC 73\\
190.6721 & 2.7743 &  13 & 1.617E+37 & -0.16 &  0.44 &  5.62 & 2    \\
190.7891 & 2.7349 &  18 & 1.978E+37 &  0.11 &  0.33 &  5.64 & 2  V\\
190.7840 & 2.6313 &  12 & 1.490E+37 &  0.15 &  0.14 &  5.69 & 2  GC 57\\
190.7018 & 2.7829 &  28 & 3.237E+37 &  0.10 &  1.00 &  5.72 & 2  SV-I\\
190.6351 & 2.7499 &  13 & 1.646E+37 & -0.15 & -0.22 &  5.74 & 2  SV-I\\
190.6443 & 2.7595 &  39 & 4.901E+37 &  0.06 &  0.56 &  5.75 & 2  GC 63\\
190.7404 & 2.7782 &  75 & 8.198E+37 &  0.05 &  0.36 &  5.77 & 2  V\\
190.8047 & 2.6861 &  17 & 1.983E+37 & -0.49 &  0.01 &  5.82 & 2    \\
190.6633 & 2.7747 &  15 & 1.907E+37 &  0.19 &  0.25 &  5.85 & 2    \\
190.7145 & 2.5904 &  23 & 3.006E+37 & -0.06 &  0.37 &  5.86 & 3  SV-S\\
190.6752 & 2.7802 &  19 & 2.284E+37 &  0.19 &  0.01 &  5.88 & 2  V,GC 45\\
190.6507 & 2.6077 &  26 & 3.574E+37 &  0.02 &  0.44 &  5.90 & 2  V\\
190.7191 & 2.7864 &  39 & 4.383E+37 &  0.14 &  0.44 &  5.96 & 3  SV-S
\\
\hline
190.7946 & 2.6365 &  57 & 6.682E+37 &  0.07 &  0.52 &  6.05 & 2  GC 70\\
190.6769 & 2.7845 &  54 & 6.464E+37 & -0.25 &  0.05 &  6.09 & 3    \\
190.6586 & 2.7768 &  49 & 5.974E+37 & -0.70 & -0.05 &  6.10 & 3    \\
190.6697 & 2.7822 &  15 & 1.843E+37 & -0.04 & -0.03 &  6.11 & 2    \\
190.7817 & 2.6162 &  15 & 1.853E+37 &  0.08 & -0.02 &  6.18 & 2    \\
190.8103 & 2.6704 &  97 & 1.108E+38 &  0.18 &  0.36 &  6.24 & 2    \\
190.6988 & 2.5840 &  58 & 7.554E+37 &  0.03 &  0.32 &  6.25 & 3  SV-S SV-I\\
190.6505 & 2.6007 &  14 & 2.724E+37 &  0.35 & -0.07 &  6.25 & 3  MV,GC
27\\
190.7200 & 2.7920 & 208 & 2.355E+38 & -0.03 &  0.26 &  6.29 & 3  GC 10\\
190.7666 & 2.7751 &  25 & 2.813E+37 & -0.06 &  0.28 &  6.32 & 3    \\
190.6517 & 2.7771 & 152 & 1.885E+38 & -0.60 & -0.01 &  6.33 & 2    \\
190.7751 & 2.7704 &  87 & 9.453E+37 &  0.14 &  0.66 &  6.39 & 3    \\
190.6832 & 2.7927 &  23 & 2.749E+37 &  0.00 & -0.11 &  6.47 & 2    \\
190.6098 & 2.6426 &  45 & 6.378E+37 &  0.10 &  0.23 &  6.47 & 2    \\
190.7157 & 2.7963 &  18 & 2.057E+37 & -0.06 & -0.33 &  6.53 & 2    \\
190.8140 & 2.7261 &  82 & 9.271E+37 &  0.01 &  0.28 &  6.78 & 2  GC 39\\
190.7129 & 2.8007 & 207 & 2.398E+38 &  0.03 &  0.31 &  6.78 & 3  V, SV-S \\
190.7639 & 2.5887 &  16 & 2.085E+37 &  0.02 &  0.40 &  6.84 & 2    \\
190.7948 & 2.7635 & 146 & 1.631E+38 &  0.06 &  0.34 &  6.92 & 3    \\
190.7406 & 2.5758 &  16 & 2.072E+37 &  0.08 &  0.37 &  7.00 & 2  SV-I\\
190.7421 & 2.7996 &  39 & 4.434E+37 &  0.19 &  0.08 &  7.02 & 3  GC 4\\
190.7293 & 2.5708 &  16 & 5.387E+37 & -0.23 &  0.23 &  7.14 & 1  V   \\
190.8258 & 2.7028 &  42 & 4.910E+37 &  0.10 &  0.62 &  7.14 & 2  GC 49\\
190.6624 & 2.5769 & 235 & 7.556E+38 & -0.79 & -0.10 &  7.19 & 1  SV-S \\
190.7416 & 2.8036 &  66 & 7.575E+37 &  0.10 &  0.36 &  7.24 & 2  V   \\
190.7074 & 2.8148 &  21 & 2.517E+37 & -0.13 & -0.31 &  7.62 & 2    \\
190.7636 & 2.5699 &  21 & 2.851E+37 &  0.14 &  0.40 &  7.83 & 2  SV-I\\
190.8375 & 2.7038 &  49 & 5.827E+37 &  0.02 &  0.60 &  7.84 & 3    \\
190.7223 & 2.8182 & 118 & 1.403E+38 &  0.10 &  0.27 &  7.87 & 3  V,SV-S \\
190.7454 & 2.8145 &  24 & 2.869E+37 &  0.26 &  0.48 &  7.93 & 2    \\
190.7236 & 2.5557 &  11 & 3.692E+37 &  0.28 &  0.67 &  7.98 & 1  SV-S \\
190.8043 & 2.7805 &  38 & 4.426E+37 & -0.04 &  0.35 &  8.03 & 2  V,GC 74\\
190.7673 & 2.8081 &  39 & 1.181E+38 &  0.48 &  0.26 &  8.05 & 1    \\
190.8414 & 2.6695 &  48 & 5.789E+37 & -0.03 &  0.36 &  8.10 & 2    \\
190.7858 & 2.7982 &  75 & 8.635E+37 &  0.14 &  0.35 &  8.12 & 2  V,MV,SV-I\\
190.8364 & 2.6455 &  36 & 4.456E+37 &  0.29 &  0.53 &  8.12 & 2    \\
190.7984 & 2.7891 &  38 & 4.447E+37 &  0.09 &  0.76 &  8.16 & 2  GC 50\\
190.7698 & 2.8098 & 101 & 1.170E+38 &  0.01 &  0.26 &  8.21 & 3    \\
190.6817 & 2.8229 &  28 & 3.510E+37 &  0.15 &  0.95 &  8.26 & 2    \\
190.8465 & 2.6912 &  35 & 4.275E+37 &  0.04 &  0.53 &  8.33 & 2  MV,SV-I\\
190.8180 & 2.6006 &  23 & 3.028E+37 &  0.16 &  1.01 &  8.43 & 2    \\
190.7072 & 2.5458 &  98 & 3.270E+38 & -0.01 &  0.09 &  8.52 & 1  SV-S \\
190.8513 & 2.6851 &  53 & 6.422E+37 & -0.04 & -0.16 &  8.61 & 2  V\\
190.8526 & 2.6939 &  68 & 8.306E+37 &  0.16 &  0.42 &  8.70 & 2  V \\
190.7123 & 2.8328 &  53 & 6.580E+37 &  0.10 &  0.57 &  8.70 & 3  SV-S \\
190.7614 & 2.8228 & 162 & 1.924E+38 &  0.07 &  0.31 &  8.72 & 3    \\
190.6728 & 2.8292 &  93 & 1.201E+38 &  0.10 &  0.59 &  8.74 & 3  SV-S \\
190.8028 & 2.5764 &  35 & 1.305E+38 &  0.09 &  0.63 &  8.79 & 3  V   \\
190.7301 & 2.8325 & 452 & 5.500E+38 &  0.00 &  0.18 &  8.79 & 3  SV-S \\
190.7847 & 2.5615 & 119 & 1.611E+38 &  0.02 &  0.33 &  8.87 & 2
MV,SV-I,GC 47\\
190.6504 & 2.8267 &  31 & 4.178E+37 &  0.08 &  0.35 &  9.01 & 2  SV-I\\
190.7924 & 2.5636 &  23 & 3.164E+37 & -0.29 &  0.47 &  9.02 & 2  V\\
190.7341 & 2.5390 & 176 & 6.143E+38 & -0.05 &  0.21 &  9.07 & 1    \\
190.7491 & 2.8339 &  74 & 8.996E+37 &  0.07 &  0.52 &  9.11 & 2  V,GC 75\\
190.8599 & 2.7032 &  35 & 4.372E+37 & -0.08 &  0.25 &  9.17 & 3    \\
190.8532 & 2.7455 & 383 & 4.630E+38 & -0.04 &  0.24 &  9.39 & 3  V,GC 38\\
190.6856 & 2.8437 &  34 & 4.476E+37 &  0.18 &  0.40 &  9.45 & 2    \\
190.7259 & 2.8444 &  23 & 2.917E+37 & -0.39 & -0.24 &  9.46 & 2  GC 71\\
190.8321 & 2.5899 &  40 & 5.338E+37 &  0.15 &  0.55 &  9.50 & 3  V \\
190.8461 & 2.7666 &  11 & 1.365E+37 &  0.22 &  0.82 &  9.55 & 2    \\
190.8317 & 2.7905 &  55 & 6.692E+37 &  0.16 &  0.49 &  9.66 & 2  \\
190.8581 & 2.6283 &  12 & 4.567E+37 & -0.08 &  0.48 &  9.70 & 1    \\
190.6874 & 2.5266 &  12 & 4.378E+37 & -0.01 &  0.67 &  9.75 & 1    \\
190.8418 & 2.7801 &  35 & 4.272E+37 &  0.17 &  0.37 &  9.77 & 2  GC 55\\
190.7280 & 2.8495 &  88 & 1.124E+38 &  0.00 &  1.52 &  9.78 & 2    \\
190.7896 & 2.5464 &  34 & 1.304E+38 &  0.04 &  0.11 &  9.80 & 1    \\
190.8519 & 2.6092 &  87 & 1.155E+38 &  0.10 &  0.39 &  9.85 & 3    \\
190.8722 & 2.6888 &  42 & 5.421E+37 & -0.13 &  0.04 &  9.87 & 2    \\
190.7901 & 2.8304 &  41 & 5.092E+37 &  0.06 &  0.40 &  9.88 & 2    \\
190.7921 & 2.5462 &  13 & 5.127E+37 & -0.32 & -0.27 &  9.89 & 1    \\
190.8692 & 2.6518 &  17 & 2.196E+37 & -0.43 &  1.06 &  9.92 & 2    \\
190.7951 & 2.8287 &  79 & 9.609E+37 &  0.06 &  0.36 &  9.95 & 3    \\
190.8739 & 2.6925 &  17 & 2.237E+37 &  0.17 &  1.16 &  9.97 & 2    \\
190.7585 & 2.8470 &  72 & 9.015E+37 &  0.11 &  0.37 & 10.03 & 2  V\\
190.7277 & 2.8547 &  84 & 1.087E+38 &  0.29 &  0.17 & 10.08 & 3  SV-S\\
190.6868 & 2.5200 &  19 & 6.576E+37 &  0.06 &  0.56 & 10.15 & 1  \\
190.6927 & 2.8563 &  37 & 5.018E+37 &  0.23 &  0.21 & 10.15 & 2    \\
190.7751 & 2.8464 &  22 & 2.859E+37 &  0.00 &  0.46 & 10.34 & 2    \\
190.8830 & 2.6974 &  13 & 4.634E+37 & -0.73 & -0.13 & 10.53 & 1    \\
190.8663 & 2.6102 &  43 & 5.825E+37 & -0.10 & -0.13 & 10.59 & 2  V,GC 67\\
190.7279 & 2.8654 &  26 & 3.548E+37 & -0.12 &  0.38 & 10.73 & 2  SV-I\\
190.8795 & 2.7406 & 198 & 2.536E+38 &  0.03 &  0.35 & 10.78 & 3  V,GC 9\\
190.8565 & 2.5832 &  25 & 3.559E+37 &  0.08 & -0.23 & 10.91 & 2    \\
190.8772 & 2.6210 & 239 & 3.258E+38 &  0.02 & -0.34 & 10.93 & 2  V\\
190.8600 & 2.7887 & 104 & 1.322E+38 & -0.30 &  0.76 & 10.96 & 2  V\\
190.8168 & 2.8347 & 292 & 3.754E+38 &  0.05 &  0.38 & 10.98 & 3    \\
190.5862 & 2.8261 &  51 & 6.021E+37 & -0.05 &  0.87 & 11.04 & 2    \\
190.8914 & 2.6655 &  99 & 1.327E+38 &  0.06 &  0.28 & 11.10 & 3    \\
190.5899 & 2.8305 & 119 & 1.423E+38 &  0.18 & -0.15 & 11.10 & 2  GC 64\\
190.8289 & 2.5477 &  19 & 7.484E+37 & -0.03 &  0.85 & 11.11 & 1  V \\
190.8773 & 2.6116 &  51 & 7.046E+37 & -0.21 &  0.41 & 11.16 & 2  GC 60\\
190.8604 & 2.7946 &   6 & 2.502E+37 & -0.03 &  0.78 & 11.18 & 1  \\
190.8825 & 2.6190 & 234 & 3.243E+38 & -0.47 & -0.03 & 11.27 & 3  V,SV-S,SV-I\\
190.8647 & 2.7920 &  96 & 1.241E+38 &  0.24 &  0.81 & 11.30 & 2  V   \\
190.8721 & 2.5928 &  11 & 1.638E+37 &  0.10 &  1.21 & 11.39 & 2  SV-I\\
190.6422 & 2.5087 & 102 & 3.627E+38 & -0.55 & -0.06 & 11.45 & 1    \\
190.8904 & 2.6316 &  21 & 7.836E+37 &  0.30 &  0.63 & 11.46 & 1  V   \\
190.8589 & 2.5708 &  26 & 3.756E+37 & -0.34 &  0.59 & 11.47 & 2    \\
190.8991 & 2.6925 &  10 & 3.637E+37 & -1.11 &  0.00 & 11.48 & 1    \\
190.9016 & 2.6698 &  25 & 3.475E+37 & -1.14 & -0.05 & 11.68 & 2  SV-I\\
190.6157 & 2.5160 & 202 & 7.260E+38 & -1.08 &  0.00 & 11.69 & 1    \\
190.8678 & 2.8009 &  56 & 7.407E+37 &  0.18 &  0.07 & 11.76 & 2    \\
190.6188 & 2.5121 & 191 & 6.864E+38 &  0.09 &  0.34 & 11.82 & 1    \\
190.8584 & 2.5550 &  48 & 7.187E+37 & -0.02 &  0.61 & 12.05 & 2    \\
190.5944 & 2.8600 & 133 & 1.597E+38 & -0.01 &  0.44 & 12.37 & 2  GC 48\\
190.8530 & 2.5411 &  14 & 5.517E+37 &  0.19 & -0.44 & 12.39 & 1    \\
190.9025 & 2.7590 &  14 & 5.503E+37 & -1.12 &  0.00 & 12.44 & 1    \\
190.8936 & 2.5930 &  20 & 7.532E+37 & -0.14 &  0.53 & 12.52 & 1    \\
190.9175 & 2.6702 &  10 & 3.726E+37 & -0.02 & -0.38 & 12.63 & 1    \\
190.8949 & 2.7845 &  11 & 4.428E+37 &  0.57 & -0.57 & 12.64 & 1    \\
190.8860 & 2.8017 &  19 & 2.666E+37 &  0.13 &  0.89 & 12.69 & 2    \\
190.9199 & 2.7405 &  26 & 9.712E+37 & -0.81 &  0.20 & 13.11 & 1  V\\
190.9021 & 2.7936 &  68 & 9.648E+37 &  0.11 &  0.44 & 13.28 & 2  V\\
190.8952 & 2.5657 &  64 & 2.467E+38 &  0.03 &  0.62 & 13.42 & 1    \\
190.9166 & 2.7686 &  24 & 9.205E+37 & -0.17 &  1.07 & 13.44 & 1  V\\
190.9256 & 2.6188 &  71 & 2.715E+38 &  0.09 &  0.55 & 13.71 & 1    \\
190.6006 & 2.8907 &  74 & 9.099E+37 &  0.32 &  0.41 & 13.77 & 2    \\
190.8863 & 2.5427 &  12 & 4.759E+37 & -0.16 &  1.76 & 13.81 & 1    \\
190.9240 & 2.7673 &  19 & 7.399E+37 &  0.14 &  0.91 & 13.83 & 1    \\
190.9210 & 2.5343 &  59 & 2.395E+38 &  0.06 &  0.54 & 15.77 & 1    \\

\enddata
\tablenotetext{a}{Net counts and luminosity (erg s$^{-1}$) are in the
  0.5 -- 2 keV band.}
\tablenotetext{b}{Soft color is defined as (M-S)/(S+M+H) where S, M,
  and H are the fluxes in bands 0.5 -- 1, 1 -- 2, and 2 -- 8 keV, respectively.}
\tablenotetext{c}{Hard color is defined as (H-M)/(S+M+H) where S, M,
  and H are the fluxes in bands 0.5 -- 1, 1 -- 2, and 2 -- 8 keV, respectively.}
\tablenotetext{d}{\textit{1}
  means the source was detected by the ACIS-S, \textit{2} means the
  source was detected by the ACIS-I, and \textit{3} means the source
  was detected by both instruments. \textit{V} denotes a long-term
  (i.e. between observations) variable and \textit{SV} denotes a
  short-term (i.e. during a single observation, either the ACIS-S
  (\textit{S}) or coadded ACIS-I (\textit{I})) variable. \textit{MV}
  indicates that the source flux varies significantly between the sequential ACIS-I observations. \textit{GC}
  indicates that the sources is matched with a globular cluster.  The
  number following \textit{GC} gives the source's position in Table \ref{t:gcmatchlst}.}
\end{deluxetable}

\clearpage
\pagebreak

\LongTables
\begin{deluxetable}{lllllllll}
\tablecolumns{9}
\tabletypesize{\footnotesize}
\tablecaption{X-ray point sources matched with globular clusters candidates in NGC 4636
\label{t:gcmatchlst}}
\tablewidth{0pt}
\tablehead{ \colhead{N} &\colhead{R.A. (X-ray)} & \colhead{Dec (X-ray)} & \colhead{Offset($^{\prime\prime}$)} & \colhead{CTIO ID\tablenotemark{a}} &
\colhead{R.A. (optical)} & \colhead{Dec (optical)} & \colhead{C-R color} & \colhead{R magnitude}  }
\startdata
 1 & 190.7004 & 2.7646 & 0.86 &  1065 & 190.70016 & 2.76461 &  1.33 & 22.98 \\
 2 & 190.6769 & 2.7526 & 0.08 &   953 & 190.67691 & 2.75253 &  1.51 & 22.23 \\
 3 & 190.7712 & 2.6908 & 0.92 &  1397 & 190.77124 & 2.69050 &  1.73 & 21.25 \\
 4 & 190.7421 & 2.7996 & 0.96 &  6947 & 190.74225 & 2.79939 &  1.13 & 22.29 \\
 5 & 190.6569 & 2.6927 & 0.48 & 13398 & 190.65686 & 2.69261 &  1.89 & 21.99 \\
 6 & 190.7647 & 2.6900 & 0.66 &  1372 & 190.76462 & 2.68983 &  0.53 & 20.67 \\
 7 & 190.7625 & 2.6541 & 0.55 &  1369 & 190.76241 & 2.65403 &  1.31 & 22.15 \\
 8 & 190.6601 & 2.6978 & 0.13 &  5431 & 190.66005 & 2.69781 &  1.55 & 20.95 \\
 9 & 190.8795 & 2.7406 & 1.46 &  8963 & 190.87991 & 2.74064 &  1.18 & 22.45 \\
10 & 190.7200 & 2.7920 & 0.16 &  1150 & 190.72008 & 2.79194 &  0.50 & 19.63 \\
11 & 190.6925 & 2.7648 & 0.13 &  1033 & 190.69255 & 2.76486 &  1.50 & 20.91 \\
12 & 190.6787 & 2.6805 & 0.36 &  5720 & 190.67862 & 2.68039 &  1.01 & 22.72 \\
13 & 190.7211 & 2.7123 & 0.46 & 11565 & 190.72099 & 2.71228 &  1.95 & 21.26 \\
14 & 190.7068 & 2.7288 & 0.29 & 11489 & 190.70676 & 2.72878 &  2.01 & 20.79 \\
15 & 190.7320 & 2.6479 & 0.65 &  6764 & 190.73192 & 2.64772 &  1.17 & 20.76 \\
16 & 190.7145 & 2.6435 & 0.60 &  6444 & 190.71432 & 2.64339 &  1.94 & 23.02 \\
17 & 190.7647 & 2.6847 & 0.76 & 15298 & 190.76462 & 2.68450 &  1.77 & 22.85 \\
18 & 190.6538 & 2.6703 & 1.19 &  5310 & 190.65363 & 2.67006 &  1.09 & 22.26 \\
19 & 190.7047 & 2.6796 & 0.79 & 13675 & 190.70488 & 2.67981 &  0.94 & 20.65 \\
20 & 190.7369 & 2.6863 & 0.51 & 13857 & 190.73683 & 2.68619 &  1.57 & 20.87 \\
21 & 190.7194 & 2.7107 & 0.53 &  6526 & 190.71925 & 2.71069 &  0.68 & 23.13 \\
22 & 190.6787 & 2.6817 & 0.33 & 13510 & 190.67859 & 2.68175 &  1.02 & 23.46 \\
23 & 190.6894 & 2.7434 & 0.24 &  5961 & 190.68933 & 2.74331 &  1.50 & 22.94 \\
24 & 190.7195 & 2.6854 & 0.96 &  6525 & 190.71921 & 2.68525 &  1.55 & 20.41 \\
25 & 190.6809 & 2.6417 & 0.33 &  5760 & 190.68088 & 2.64158 &  1.38 & 21.06 \\
26 & 190.7296 & 2.6897 & 0.27 & 13824 & 190.72963 & 2.68958 &  1.45 & 23.36 \\
27 & 190.6505 & 2.6007 & 0.87 & 15478 & 190.65050 & 2.60050 &  0.31 & 23.04 \\
28 & 190.6781 & 2.7048 & 0.55 &   960 & 190.67825 & 2.70486 &  1.82 & 20.39 \\
29 & 190.7056 & 2.7652 & 0.25 &  6281 & 190.70566 & 2.76522 &  1.41 & 21.28 \\
30 & 190.7080 & 2.6068 & 1.08 &  1098 & 190.70799 & 2.60650 &  1.79 & 21.20 \\
31 & 190.6797 & 2.7138 & 0.14 &  5739 & 190.67963 & 2.71386 &  1.45 & 20.64 \\
32 & 190.7348 & 2.6449 & 0.66 & 13847 & 190.73462 & 2.64481 &  1.75 & 20.97 \\
33 & 190.6708 & 2.6102 & 1.16 & 13483 & 190.67082 & 2.60986 &  0.02 & 22.70 \\
34 & 190.7646 & 2.6689 & 0.97 &  7355 & 190.76457 & 2.66864 &  1.66 & 22.89 \\
35 & 190.6779 & 2.7223 & 0.17 &  5705 & 190.67783 & 2.72228 &  1.35 & 22.94 \\
36 & 190.7556 & 2.6970 & 0.64 &  7203 & 190.75549 & 2.69692 &  1.34 & 21.79 \\
37 & 190.6907 & 2.7206 & 1.19 &  5983 & 190.69046 & 2.72042 &  1.88 & 21.07 \\
38 & 190.8532 & 2.7455 & 0.60 &  1707 & 190.85333 & 2.74550 &  0.75 & 20.30 \\
39 & 190.8140 & 2.7261 & 0.68 &  8118 & 190.81387 & 2.72597 &  1.30 & 22.18 \\
40 & 190.6447 & 2.7308 & 0.23 &  5145 & 190.64467 & 2.73078 &  1.39 & 21.20 \\
41 & 190.6289 & 2.6775 & 0.46 &  4905 & 190.62888 & 2.67739 &  1.81 & 21.81 \\
42 & 190.7231 & 2.6155 & 0.60 & 11574 & 190.72308 & 2.61533 &  1.75 & 21.29 \\
43 & 190.7668 & 2.7411 & 1.50 &  7388 & 190.76653 & 2.74081 &  1.66 & 20.39 \\
44 & 190.7333 & 2.7030 & 0.84 & 11643 & 190.73305 & 2.70289 &  1.71 & 23.14 \\
45 & 190.6752 & 2.7802 & 0.05 &  5651 & 190.67525 & 2.78025 &  1.61 & 20.24 \\
46 & 190.7250 & 2.6979 & 0.65 &  6619 & 190.72482 & 2.69783 &  1.66 & 21.46 \\
47 & 190.7847 & 2.5615 & 0.96 &  7666 & 190.78458 & 2.56131 &  0.03 & 22.12 \\
48 & 190.5944 & 2.8600 & 1.26 &  4352 & 190.59471 & 2.85992 &  0.11 & 22.10 \\
49 & 190.8258 & 2.7028 & 0.91 &  8282 & 190.82558 & 2.70283 &  0.20 & 21.40 \\
50 & 190.7984 & 2.7891 & 1.03 &  7866 & 190.79807 & 2.78914 &  1.58 & 21.41 \\
51 & 190.7773 & 2.6984 & 0.84 &  7545 & 190.77711 & 2.69828 &  1.78 & 22.49 \\
52 & 190.7330 & 2.6690 & 0.72 &  6780 & 190.73286 & 2.66892 &  1.77 & 22.77 \\
53 & 190.6644 & 2.7046 & 0.56 & 11250 & 190.66429 & 2.70456 &  1.87 & 22.06 \\
54 & 190.6927 & 2.7157 & 0.74 & 13583 & 190.69255 & 2.71553 &  1.68 & 21.26 \\
55 & 190.8418 & 2.7801 & 1.30 &  1655 & 190.84216 & 2.78000 &  1.07 & 19.72 \\
56 & 190.7938 & 2.7207 & 1.00 &  1482 & 190.79353 & 2.72056 &  1.76 & 22.32 \\
57 & 190.7840 & 2.6313 & 0.47 &  7652 & 190.78391 & 2.63119 &  1.74 & 22.34 \\
58 & 190.6774 & 2.7211 & 0.57 &  5692 & 190.67725 & 2.72097 &  0.14 & 21.26 \\
59 & 190.7496 & 2.6800 & 0.41 &  7081 & 190.74950 & 2.67989 &  1.57 & 22.55 \\
60 & 190.8773 & 2.6116 & 0.93 &  8933 & 190.87746 & 2.61133 &  0.93 & 22.66 \\
61 & 190.6699 & 2.6470 & 0.22 &   918 & 190.66991 & 2.64692 &  1.27 & 21.70 \\
62 & 190.6877 & 2.6931 & 0.55 & 13554 & 190.68767 & 2.69294 &  1.29 & 22.87 \\
63 & 190.6443 & 2.7595 & 0.75 &  5133 & 190.64412 & 2.75936 &  1.63 & 20.70 \\
64 & 190.5899 & 2.8305 & 0.21 & 10851 & 190.58997 & 2.83044 &  0.92 & 21.81 \\
65 & 190.7268 & 2.7077 & 0.71 &  6655 & 190.72670 & 2.70756 &  1.08 & 21.77 \\
66 & 190.7796 & 2.6748 & 0.86 &  1435 & 190.77940 & 2.67458 &  1.50 & 21.37 \\
67 & 190.8663 & 2.6102 & 0.32 &  8799 & 190.86624 & 2.61025 &  0.23 & 22.20 \\
68 & 190.7299 & 2.7745 & 0.63 &  6707 & 190.72980 & 2.77442 &  1.86 & 21.90 \\
69 & 190.6229 & 2.7059 & 0.66 &  4797 & 190.62283 & 2.70572 &  0.10 & 22.53 \\
70 & 190.7946 & 2.6365 & 0.61 &  7822 & 190.79445 & 2.63642 &  1.74 & 21.86 \\
71 & 190.7259 & 2.8444 & 0.71 & 11588 & 190.72574 & 2.84436 &  1.17 & 20.44 \\
72 & 190.7317 & 2.5883 & 0.76 &  1215 & 190.73149 & 2.58814 &  1.61 & 20.06 \\
73 & 190.7851 & 2.6359 & 0.66 &  7670 & 190.78496 & 2.63578 &  1.85 & 23.09 \\
74 & 190.8043 & 2.7805 & 0.84 &  7973 & 190.80437 & 2.78031 &  0.02 & 22.51 \\
75 & 190.7491 & 2.8339 & 0.71 & 13910 & 190.74892 & 2.83386 &  1.36 & 23.23 \\
76 & 190.6942 & 2.7729 & 0.70 &  6052 & 190.69403 & 2.77278 &  1.73 & 21.57 \\
77 & 190.7134 & 2.6774 & 1.01 & 15522 & 190.71320 & 2.67719 &  0.92 & 23.01 \\
\enddata
\tablenotetext{a}{Dirsch et al. 2005}
\end{deluxetable}

\end{document}